\documentclass[pdftex,%
	usenatbib,usegraphicx]{mn2e}
\usepackage{times}

\usepackage[utf8]{inputenc}
\usepackage[UKenglish]{babel}

\usepackage[fleqn]{amsmath}
\usepackage{aas_macros}
\usepackage{bbold}

\usepackage[pdftex,%
	bookmarks,%
	plainpages=false,%
	unicode=true,%
	pdfauthor={Cornelius Weig and Torsten Ensslin},%
	pdftitle={Bayesian analysis of spatially distorted cosmic signals from Poissonian data},%
	colorlinks=false]{hyperref}

\DeclareMathOperator{\plog}{W}
\DeclareMathOperator{\res}{\mathrm{Res}}

\providecommand{\figref}[1]{Fig.~\ref{#1}}
\providecommand{\tableref}[1]{\tablename~\ref{#1}}
\providecommand{\openone}{\mathbb{1}}
\newcommand{\e}{\mathrm{e}}

\newcommand{\NormWithMetric}[2]{#1^\tp #2 #1}
\newcommand{\Lpnorm}[2]{\lVert#1\rVert_{#2}}
\renewcommand{\det}[1]{|#1|}
\newcommand{\gauss}[2]{\frac{1}{\det{2\pi\,#2}^{1/2}}\,\e^{-\NormWithMetric{#1}{#2^{-1}}/2}}
\newcommand{\gaussSymbol}[2]{\mathcal{G}(#1,\,#2)}

\newcommand{\prop}{\ensuremath{\mathbf{S}}}
\newcommand{\rprop}{\ensuremath{\mathbf{T}}}
\newcommand{\propD}{\mathbf{D}}
\newcommand{\mixingMatrix}{\ensuremath{\mathbf{R}}}
\newcommand{\strue}{\ensuremath{s_t}}

\newcommand{\Ptilde}{\widetilde{P}}
\newcommand{\PathInt}[1]{\int\!\mathcal{D}#1}
\newcommand{\expectationWRT}[2]{\left\langle#1\right\rangle_{#2}}
\newcommand{\tp}{\mathtt{t}}
\newcommand{\variation}[2]{\frac{\delta #1}{\delta #2}}
\newcommand{\MAP}{\ensuremath{m_{cl}}}
\newcommand{\MAPu}{\ensuremath{m_{cl}^{u}}}
\newcommand{\mnaive}{\ensuremath{m_{na}}}

\def\jointprob{\ensuremath{P(s,\,d)}}
\def\wrt{with respect to}
\def\FOG{\emph{finger-of-god}}

\def\mpc{\mathrm{Mpc}}
\def\PDF{PDF}
\def\powsp{power spectrum}

\begin{document}
\title[Reconstruction of spatially distorted fields]%
	{Bayesian analysis of spatially distorted cosmic signals from Poissonian data}
\author[Cornelius Weig and Torsten A.~Enßlin]{Cornelius Weig$^1$ and Torsten A.~Enßlin$^1$\\
$^1$ MPA, Max-Planck Institut für Astrophysik, Karl-Schwarzschildstr.~1, D-85748 Garching, Germany
}

\maketitle

\begin{abstract}
Reconstructing the matter density field from galaxy counts is a problem
frequently addressed in current literature.
Two main sources of error are shot noise from galaxy counts and insufficient
knowledge of the correct galaxy position caused by peculiar velocities and
redshift measurement uncertainty.
Here we address the reconstruction problem of a Poissonian sampled log-normal
density field with velocity distortions in a Bayesian way via a maximum a
posteriory method.
We test our algorithm on a 1D toy case and find significant improvement
compared to simple data inversion.
In particular, we address the following problems: photometric redshifts,
mapping of extended sources in coded mask systems, real space reconstruction
from redshift space galaxy distribution and combined analysis of data with
different point spread functions.
\end{abstract}

\section{Introduction}
Correct and thorough signal analysis is vital in cosmology.
This is for one thing due to often low signal-to-noise ratios of many cosmic
measurements.
An important fact to notice here is that many measurements can \emph{not} be
independently repeated, as nature only grants us with one realisation of the
data such as the CMB or the large-scale structure of the universe (LSS).

Since the complications in extracting the desired information from data are so
fundamental, it is often not possible to draw sensible conclusions from it
without some knowledge on the properties of the underlying signals.
Although it would be desirable to be completely independent and prejudice-free,
one typically has to give up on some freedom in the analysis by restricting to
a specific model.
In return one gains much better constraints on the measured quantity.
Hence a thorough signal analysis must take care of all those aspects, otherwise
the wrong conclusions could be drawn.
This is most naturally done in the Bayesian framework where all variables are
considered to be subject to error and variance.

The emphasis in this work is on the reconstruction problem of a log-normal
density field that is sampled via a Poissonian process as simple description of
galaxy formation and a number of other processes, like a highly structured gamma
ray emissivity.
We choose the log-normal field, because we think it suited to model the dark
matter density distribution of the Universe (see section
\ref{section:log-normal-density field}).
In addition, we extend the problem such that the signal is spatially
distorted and galaxy counts from one position may show up at other locations.
This way a sharp peak in the signal can show up as a broad distribution in the
data depending on the underlying distortion process.
This allows us to address the real space reconstruction problem of the dark
matter density field from redshift distorted galaxy counts and to naturally
incorporate photometric redshift errors in our analysis, to name a few.

The most generic case of uncertain position is the measurement error from the
measurement apparatus itself which comes with \emph{any} measurement.
In many cases, these errors form a Gaussian distribution around a mean value.
But there are also other cases like photometric redshift where due to the
measurement technique there is considerable chance for `catastrophic
outliers' which leads to non-Gaussian probability distributions.
In our analysis the chance for such catastrophic errors can be naturally
included and dealt with.
This permits do deal with cases where such outliers are the rule, for example
coded mask detectors in X- and $\gamma$-ray astronomy where a point source has
to be identified via its complex mask shadow on the detector plane.
Other areas where spatial distortion should be included in the analysis are
$\gamma$-ray astronomy via \v Cerenkov telescopes, or Ultra-high-energy cosmic
ray detectors due to the extended point spread functions of the measurement
devices.

In all examples so far the distortion of the data was known a priori and was
fixed.
However, one can go one step further and allow the distortion to depend on the
signal itself that one wants to measure.
The paradigm for this problem is the measurement of redshift in galaxy surveys.
Here the aim is to measure the real space density distribution of matter in
the Universe via galaxy counts, but since the presence of matter has an effect
on the peculiar velocities of the observed galaxies, the distortion depends on
the details of the signal to be reconstructed.
Beyond the linear regime, where a dark matter halo has collapsed to a virialised
object like a galaxy cluster, the galaxies have large peculiar velocities.
Since only the component along the line of sight adds to the redshift, those
collapsed objects appear as dense elongated structures in redshift space
pointing towards the observer, therefore this is also called the `\FOG'
effect.

Including the feedback of matter on the redshift space distortions is the
ultimate goal of LSS reconstruction.
The point of our work is to address one very important effect so far often
ignored in statistical inference of the LSS: spatial distortions.
In order to focus our discussion on this, we work with simplified descriptions
of the complex galaxy formation process.
The adopted description, however, was previously shown to provide good
reconstructions despite its simplicity.

Significant progress in the field of large-scale structure reconstruction with a
log-normal model for the dark matter over-density has been achieved in a number
of recent works.
\cite{2009PhRvD..80j5005E} derive the MAP estimator and loop corrections thereof
within the \emph{information field theory} (IFT) framework.
\citet{2010MNRAS.403..589K} successfully apply the MAP Poissonian log-normal
filter on mock data from N-body simulations.
\citet{2009arXiv0911.2498J} even
achieved a reconstruction from real world SDSS\,7 data going beyond the MAP
approximation by using a Hamiltonian sampling method.
However, all these approaches do not take the spatial uncertainty of redshift
measurements -- i.e.~the point spread function (PSF) -- into account.
The complications from a non-trivial PSF has been addressed in a number of works
with similar data models as ours.
\cite{1992ITSP...40.2290H,Green90bayesianreconstructions,2008PMB....53..593W}
consider a similar data model but use a signal clique prior adequate for image
reconstruction.
\cite{2009OptCo.282.2489O} have a distorted Poissonian data model but use a
smoothing prior based on Fisher information for the signal.
\cite{1990Ap&SS.171..341N} also work with a Poissonian data model but use an
ad-hoc image entropy as prior for their signal.
\cite{1978JOSA...68...93F,1978Natur.272..686G} work with maximum entropy prior
for the signal distorted by a point spread function and approximate the
Poissonian distribution by a Gaussian.
Different choices for such entropy priors are discussed in
\cite{1982JApA....3..419N,1985A&A...143...77C}.

The outline of this work is as follows.
We formulate the Bayesian reconstruction problem in a field-theoretic language
in section \ref{section:theoretic-background}.
In section \ref{chapter:reconstruction-of-spatially-distorted-signals} we
address the reconstruction problem from spatially distorted data.
A suitable data-model for this purpose is introduced.
In particular this includes a discussion of the distortion operator in section
\ref{section:the-distortion-operator}.
We address how approximate error bars can be constructed for our
reconstructions.
In section \ref{section:photometric-redshift} we apply our method to the LSS
reconstruction from photometric redshift measurement in a 1D test-case.
We show how data sets with different error characteristics can be naturally
combined within our framework in the subsequent section.
In section \ref{section:coded-mask} we apply the same algorithm to a completely
different field, namely X- and $\gamma$-ray astronomy via coded mask
telescopes.
Finally, in section \ref{section:s-dependent-distortion} we formulate a
distortion problem where the distortion itself depends on the signal to be
reconstructed.
We therefore propose a model how to approach the forward problem to transform
from real into redshift space.
We compare our results for this distortion model to a Metropolis-Hastings
sampling method in section \ref{section:metropolis-hastings}.
In \ref{section:conclusions} we summarise our findings.
Details about the notation can be found in \ref{appendix:notation}.

\section{Theoretic framework}\label{section:theoretic-background}
\subsection{The way from data to information}
\label{section:data-to-information}
There is always a difference between the data that we measure and the
information we want to extract from the data.
The process from mere data acquisition to information extraction needs a model
for the way how data is produced from a signal $s$.
In a practical application one measures data $d$ and tries to infer the signal
that produced the data.
In most cases this data inversion is not unambiguous when e.g.~the signal $s$ is
continuous and the data $d$ are discrete.
Then, one needs to formulate the data model as probability distribution
functions (\PDF) such as $P(d|s)$, $P(s)$, $P(d)$, $P(s|d)$ which we call
-- following standard Bayesian naming conventions -- \emph{likelihood},
\emph{prior}, \emph{evidence} and \emph{posterior}, respectively.

This model may include physical and technical processes all the like, such as
noise from the measurement system or natural and unavoidable sources of error.
Noise in particular makes the mapping from signal to data ambiguous and
non-deterministic such that different signals can produce the same data.

Although it was shown that it is in principle possible to extract prior
information on the signal $s$ from the data \citep{2010arXiv1002.2928E}, we will
always assume that $P(s)$ is given in advance.
In many situations, the prior is taken to be flat with the intention to be
`prejudice-free'.
However, the flatness of $P(s)$ depends on the coordinate system chosen for $s$
and is therefore a (hidden) choice for a specific coordinate system.
Hence working with an explicit prior should not be seen as a flawed bias for a
specific model but as a \emph{complete} discussion of the problem.

\subsection{Optimal map making}\label{section:optimal-map-making}
\index{map!optimal}
Unfortunately, in most situations the process of data generation from the true
signal \strue\ is not reversible which means that some information is
irreversibly lost and \strue\ can not be fully restored from the data.
So the task of signal inference from a data set $d$ is to produce a
\emph{map} $m$ as an approximation of \strue\ provided the likelihood $P(d|s)$
how data emerges from the signal and the prior $P(s)$.
A reasonable map making strategy is to minimize the average error that one makes
in the reconstruction of \strue.
Here it is of key importance how the errors are weighted.
A suitable measure for the error is the $L_2$-norm defined as
\begin{equation}
	\Lpnorm{s}{2} \equiv
		\left(\int dx\,|s_x|^2\right)^{1/2}
	\label{fundamentals:L2-norm:def}
\end{equation}
where the appearing integral is a volume integral over the whole position space.

Minimizing the expected error $\expectationWRT{\Lpnorm{s-m}{2}^2}{(s|d)}$
provides the prescription
\begin{equation}
	m_x = \expectationWRT{s_x}{(s|d)} \equiv \PathInt{s}\,s_x\,P(s|d),
	\label{fundamentals:map}\index{map}
\end{equation}
for map making, where the subscript $x$ refers to the point or pixel whose
estimated map value is to be calculated.
Since $s$ is a field, the $\mathcal{D}s$-integral is a path integral which
runs over all possible field configurations.
Usually one has to take thorough care about the convergence of path integrals,
but in all practical applications one deals with a finite number of signal
points or pixels, such that the path integral reduces to a volume integral over
a rather high but finite dimensional space.

Of course, different $L_p$-norms lead to different map making techniques.
Thus it is a question of taste and belief which $L_p$-norm one prefers and
therefore which map making technique one considers best.
However, since we know that the reconstruction cannot fully recover the exact
signal, it makes sense to be generous to small deviations but penalise large
ones strongly, which is exactly what the $L_2$-norm does.

As a word of caution, it should be mentioned here that although
\eqref{fundamentals:map} is independent of the coordinate system chosen for
the data $d$, it does depend on the coordinates of $s$, however.
Thus we need to specify in which signal coordinates we want to minimize the
reconstruction error.

\subsection{IFT formulation of moment calculation}
Moment calculation can also be formulated in the language of statistical field
theory and since we are dealing with information here, we call this framework
\emph{information field theory}\index{IFT} (IFT).
This was addressed by e.g.~%
	\cite{2009PhRvD..80j5005E,
	1996PhRvL..77.4693B,
	1999physics..12005L,
	2001AIPC..568..425L,
	2001EPJB...20..349L,
	2005EPJB...46...41L}
but for completeness, we briefly summarise their findings.
Bayes' law allows to rewrite \eqref{fundamentals:map} as
\begin{equation}
	\expectationWRT{s_x}{(s|d)} = \PathInt{s}\,s_x \frac{P(s)\,P(d|s)}{P(d)}
	 = \PathInt{s}\,s_x \frac{\jointprob}{P(d)}
	\label{fundamentals:map-2}.
\end{equation}
As the set of all possible signals $s$ is exclusive and exhaustive, one can
marginalise \jointprob\ over $s$
\begin{equation}
	P(d)=\PathInt{s}\,\jointprob \label{fundamentals:marginalisation}.
\end{equation}
So it turns out, that the only required probability distribution for moment
calculation is \jointprob.

The crucial step now is to realise that the above integrals can also be
formulated in the language of statistical field theory
as e.g.~\citet{1999physics..12005L}, \citet{2009PhRvD..80j5005E},
\citet{PhysRev.106.620} and others proposed.
Following amongst others the idea of
\citet{1953JChPh..21.1087M,
	citeulike:1015842,
	citeulike:893708,
	1987PhLB..195..216D}
	, we define a probability Hamiltonian as
\begin{equation}
	H_d[s] \equiv -\log \jointprob \label{fundamentals:hamiltonian}
\end{equation}
which leads to the IFT equivalent of the partition function
\begin{equation}
	Z_d \equiv \PathInt{s}\, \e^{-H_d[s]} \label{fundamentals:partitionsum}.
\end{equation}
Comparing \eqref{fundamentals:partitionsum} with
\eqref{fundamentals:marginalisation} one recognizes that $P(d)=Z_d$.
So the posterior can be expressed as $P(s|d)=\e^{-H_d[s]}/Z_d$.

The full advantage of this formulation becomes evident when the posterior is or
is approximated by a Gaussian, i.e.
\begin{equation}
	P(s|d)=\gaussSymbol{s-m}{\propD}\equiv\gauss{(s-m)}{\propD}.
	\label{def:datamodel:prior}
\end{equation}
Then one can analytically calculate the \emph{generating functional}
\begin{equation}
	Z_d[J]\equiv\PathInt{s}\,\e^{-H_d[s]+J^\tp s}
		= \e^{J^\tp m + \NormWithMetric{J}{\propD}/2}
	\label{fundamentals:generating-functional-gaussian}
\end{equation}
where $J$ is an arbitrary field that we call \emph{source field} in analogy to
quantum field theory.
From $Z_d[J]$ all moments can be calculated by functional derivation:
\begin{equation}
	\expectationWRT{s_{x_1}\!\cdots s_{x_n}}{(s|d)}  = \left.\frac{1}{Z_d} \frac{\delta^nZ_d[J]}{\delta J_{x_1}\!\cdots \delta
		J_{x_n}}\right|_{J=0}.
	\label{fundamentals:moments-from-genfunc}
\end{equation}

This result will be useful in section \ref{section:errorestimation} where we
address error estimation.

\subsection{MAP approximation}\label{section:classical-map}
For many applications calculating the partition function is not feasible,
because the joint probability is too complex that the generating functional can
be calculated analytically.
In these cases one has to resort to an approximation of
$\expectationWRT{s}{(s|d)}$ such as the \emph{maximum a posteriori} (MAP) map
\MAP, which approximates the posterior mean by the map that maximises $P(s|d)$.
The evidence only serves as normalisation constant, so maximising the
posterior comes down to maximising the joint probability of signal and data.
And since \jointprob\ can be expressed in terms of $H_d[s]$, maximising
$\jointprob = \e^{-H_d[s]}$ is -- due to the monotony of the exponential
function -- equivalent to minimizing the Hamiltonian.
Minimizing the Hamiltonian is a well-known principle from classical mechanics so
it makes sense to call the MAP map the \emph{classical map}.

The posterior average $\expectationWRT{s}{(s|d)}$ and \MAP\ are exactly equal,
if $P(s-\MAP|d)$ is a symmetric and singly peaked function.
In the more interesting cases however, this is usually not the case.
Yet $\expectationWRT{s}{(s|d)}\approx\MAP$ often holds when $P(s|d)$ is sharply
peaked.
While the posterior mean minimizes the expected $L_2$-distance from
reconstruction to the true signal, one can show that the classical map minimizes
the $L_0$-distance.

\section[Reconstructing distorted signals]{Reconstruction of spatially distorted
	signals}\label{chapter:reconstruction-of-spatially-distorted-signals}
\subsection{Signals with uncertain position measurement}
The reconstruction of a signal from data with distorted or imprecise position
information is a generic problem class.

By \emph{signal}\index{signal} we mean a distributed physical quantity, i.e.~a
field.
In principle, the signal must be considered to be continuous but for most
practical applications it must be sampled, which yields a field vector $s$.
Therefore, we assume from now on that $s$ is indeed an ordinary vector from a
high dimensional real vector space.
We also call this space of all possible signals the \emph{signal phase
space}\index{signal!phase space}, its elements $s$ a \emph{signal
realisation}\index{signal!realisation} and the value of $s$ at location $i$
the \emph{field strength}\index{field strength} $s_i$.

The signal can be probed by a measurement apparatus which ultimately produces
\emph{data}\index{data} which are discrete by nature.
From this alone it is clear that data and signal are a priori defined on
different vector spaces.
In the following, the data will always be the count rates of
\emph{events}\index{event}.
Examples for these events could be the detection of galaxies in some direction
with redshift in a specific range or the registration of photons in a X-ray
detector.
The number of such events as a function of our data space coordinates then form
a data vector.

The \emph{response}\index{response} $\mu$ of an experimental set-up is by
definition the expectation value of the data $d$ averaged over all possible data
realisations:
\begin{equation}
	\mu\equiv \expectationWRT{d}{(d|s)}
	\label{def:response}
\end{equation}
In other words, the response is perfect \emph{noiseless}\index{noise!noiseless}
data.
We say that the response is \emph{local}\index{local}, when the field strength
$s_i$ triggers only events in one data bin $d_{j}$.
If on the other hand $s_i$ may trigger events in a number of data bins
$d_{j_1},\,d_{j_2}\ldots$ we say that the response is
\emph{non-local}\index{non-local}.
If in addition the same data bin $d_j$ may be filled by different
$s_{i_1},\enspace s_{i_2}$ we call the data space
\emph{distorted}\index{distortion}.
We address both the problems where the distortion is independent of the
underlying signal but also the case where the distortion does depend on the
signal.

In this work we always assume that the signal has a log-normal \PDF\ with known
covariance.
We choose this signal, since we believe that it approximately models the
large-scale matter distribution and other signals.

\subsubsection{The log-normal distribution for matter}%
	\label{section:log-normal-density field}
There are several good reasons why to believe that the log-normal \PDF\ models
the large-scale matter distribution well.
For one thing it has already been used by \citet{1934ApJ....79....8H} as early
as \citeyear{1934ApJ....79....8H}, to successfully model the galaxy count rates
in 2D sky patches.
For another \citet{1991MNRAS.248....1C} found that if an initially Gaussian
random field, as it is predicted by most inflationary models and more
importantly as it is observed in the CMB, evolves over time and the peculiar
velocities grow linearly, then initial Gaussian field is evolved over time into
a log-normal field.

There is also evidence from N-body simulations that the log-normal \PDF\ is an
adequate description of the large scale matter distribution.
\citet{2001ApJ...561...22K} found by direct comparison of the one-point and
two-point correlation functions obtained from N-body simulations to the
one-point and two-point log-normal \PDF\ that the former can be very accurately
modelled by the latter even in the strongly non-linear regime with
overdensities up to 100.
\citet{2010MNRAS.403..589K} could even show with mock tests that their
Poissonian log-normal model was able to reconstruct the matter distribution
accurately for overdensities up to 1000.
Furthermore, \citet{2009ApJ...698L..90N} show also with data from N-body
simulations that a log-normal density field fits the density \powsp\ much better
in the slightly non-linear regime than Gaussian fields do.
In both of the above works, the N-body simulations were seeded with small
Gaussian density fluctuations.
Recently, the log-normal model has also been successfully applied to matter
density reconstruction from SDSS data
\citep{2009arXiv0911.2498J,
2009arXiv0911.2496J}.

And last but not least: the log-normal \PDF\ is mathematically simple and
therefore comparatively easy to handle.

Hence we assume that the matter density $\rho$ in the universe can be modelled
by
\begin{equation}
	\rho \equiv \rho_0\,\e^s.
	\label{def:datamodel:rho}
\end{equation}
Here $\rho_0$ is a reference density, close to but different from the mean
density, the exponential function is meant to be taken component-wise and the
signal $s$ shall be a field with Gaussian\index{Gaussian \PDF} covariance
matrix \prop, i.e.~$P(s)=\gaussSymbol{s}{\prop}$.
Throughout this work we will always assume that $s$ is a Gaussian field with
covariance matrix \prop, if not stated otherwise.

One characteristic of the log-normal distribution is that it generates very
large overdensities in the case of large \prop, while the inner structure of
void regions is barely visible to the eye.
Besides, the log-density field contains information about the primordial density
fluctuations.
Therefore, instead of reconstructing the density field itself, we reconstruct
the log-density field $s=\log\bigl(\rho/\rho_0\bigr)$ which is Gaussian for the
log-normal distribution.

\subsubsection{The data model}\label{section:data-model}
\paragraph{The local response:}
As argued in section \ref{section:data-to-information}, the data model must
provide a likelihood to obtain some data $d$ given some signal $s$.
Since only a minor portion of matter radiates, we have to rely on tracers of
matter density.
It is widely accepted that galaxy count rates can serve as tracers for the
\emph{dark matter}\index{dark matter} density field on large scales.

As a first step we have to find a formulation for the local response to the
signal.
Therefore, we subdivide the observed space into boxes of volume $V_i$ and
relate $\mu_i$, the expected number of events within $V_i$, to the field
strength $s_i$, the signal strength averaged over $V_i$.
We assume the \emph{local response}\index{response!local} $\mu$ of the
observation to be given by
\begin{equation}
	\mu_i[s] = w_i\, V_i \, \rho_e^{(0)} \biggl(\frac{\rho_i}{\rho_0}\biggr)^b = \kappa_i \, \e^{b\,s_i}.
	\label{def:datamodel:mu-local}
\end{equation}
Here $w$ is the window function which encodes information on the exposure time
and survey geometry that might have an impact on the detection probability
in $V_i$ and $\rho_e^{(0)}$ is some reference event density.
For brevity we have defined $\kappa_i = w_i\,V_i\,\rho_e^{(0)}$ which we call
\emph{zero-response}\index{response!zero-response}, because $\mu[s=0] = \kappa$.
Throughout this work we always assume that $\kappa$ is known a priori.

The bias $b$ determines, how fast and how strong overdensities produce events.
There are strong reasons to assume that a single linear bias factor is an
oversimplification of nature as
e.g.~\cite{1993ApJ...413..447F,
2008PhRvD..78h3519M,
2008PhRvD..78j9901M,
2009ApJ...703.1230J} 
show.
Nevertheless, for a proof-of-concept at which we aim the single bias factor
simplifies the set-up and preserves the relevant features of more elaborate
models.
Besides, a scale-dependent bias for $s$ can be incorporated in the covariance
matrix by working in the Fourier picture \citep{2010MNRAS.403..589K}.
For $b \ll 1$ the response can be expanded as a power series in $s$
\begin{equation*}
	\mu_i[s] = \kappa_i\sum_{j=0}^\infty \frac{(b\,s_i)^j}{j!}\approx
		\kappa_i(1+b\,s_i)
\end{equation*}
which is the familiar form of a linear bias model used in galaxy cosmology.
However, when $b$ approaches unity for a signal with $O(1)$ variance, higher
orders of $s$ become more and more important and the linear approximation is
bound to fail.

It must be stressed, that $\rho_e^{(0)}$ is not the average event density,
because $\bar{\rho}_{e,i}\equiv \expectationWRT{\mu_i}{(s)}= \kappa_i
\PathInt{s}\, \e^{b\,s_i}\gaussSymbol{s}{\prop} = \kappa_i\,\e^{b^2\prop_{ii}/2}$
\citep[e.g.~][]{1991MNRAS.248....1C,
	2001ApJ...561...22K
	}.
When setting up a simulation one usually specifies $\bar{\rho}_e$ and not
$\rho_e^{(0)}$, so one has to keep this relation in mind when the bias is
varied.

\paragraph{The Poissonian sampling process:}%
	\label{section:poisson-sampling-process}
So far, we have only defined the response, which specifies how many events one
can expect on average in $V_i$ provided the signal $s$.
But since the number of events in $V_i$ is always a (non-negative) natural
number, one gets the true response only if the events are averaged over many
different data realisations.
However, in many cases nature prohibits this practice, because the number of
events $d_i$ is a random number with expectation value $\mu_i$ that is not
redrawn between observations, such as the random number of galaxies in $V_i$.
Therefore, the noise inflicted by the inaccurate approximation of the response
by the events must be included in the data model.

The Poissonian \PDF\ describes the number of events when the number of possible
outcomes of an experiment are vast, but only a few counts are expected.
This applies to the case for the observed number of galaxies in a box $V$
where we have only the information of the expected number of enclosed galaxies.
Of course it would be desirable to include more knowledge about the environment
of a galaxy into the \PDF\ for their abundance as many semi-analytic models do
\citep[e.g.~][]{%
	1991ApJ...379...52W,
	1993MNRAS.264..201K,
	2000MNRAS.317..697E,
	2000MNRAS.319..168C}. 
Many N-body simulations indicate that there are numerous parameters that could
or should be taken into account \citep[e.g.~][]{
	1996ApJ...462..563N,
	2005Natur.435..629S,
	2001MNRAS.321..372J}. 
However, adding more complexity makes the problem even harder to tackle than the
Poissonian \PDF.
Besides, due to the generic nature of the Poissonian \PDF, the method can also
be applied to completely different problem class, such as the shot noise of
X-ray photons in a detector.

We now make the assumption, that $d_i$, the number of events in $V_i$, is
independent in its noise properties from all other data bins.
In these assumptions we follow the works of \citet{%
	1956AJ.....61..383L,
	1980lssu.book.....P,
	2002sgd..book.....M,
	2009MNRAS.400..183K}.

Thus the likelihood of the data becomes
\begin{equation}
	P(d|\mu) = \prod_i \frac{\mu_i^{d_i}\,\e^{-\mu_i}}{d_i!}.
	\label{def:datamodel:likelihood}\index{likelihood}
\end{equation}
As $\mu$ can be expressed in terms of $s$, this provides the desired likelihood
$P(d|s)$.
The data $d$ can also be viewed upon as the Poissonian sampled response.
While $\mu$ itself is by definition noiseless, every realisation $d$ has
\emph{Poissonian noise}\index{noise!Poissonian}.

\paragraph{The distortion operator:}\label{section:the-distortion-operator}
\index{distortion!operator}
So far, the response \eqref{def:datamodel:mu-local} is entirely local.
We now propose the concept of a distortion operator which naturally introduces
non-locality in the response \eqref{def:datamodel:mu-local}.

Since it is the data space that is distorted, one should apply the distortion to
the local response \eqref{def:datamodel:mu-local}.
Hence we define the distorted response
\begin{equation}
	\mu_i = \sum_j \mixingMatrix_{ij}\kappa_j\,\e^{b\,s_j},
	\label{def:datamodel:mu-general}\index{distortion!matrix}
\end{equation}
with a distortion matrix \mixingMatrix.

It is a reasonable requirement that the distortion matrix should preserve the
expected number of events, i.e.
\begin{equation}
	\sum_i \mu_i  = \sum_i \kappa_i \e^{b\,s_i}.
	\label{datamodel:conservation-of-mu}
\end{equation}
The most natural explanation of \mixingMatrix\ is to interpret
$\mixingMatrix_{ij}$ as the probability that an event which occurred at position
$x_j$ in signal space is observed at location $x_i$ in data space and increases
$d_i$.
This is also how we set up \mixingMatrix\ in practice and it is easy to show
that a distortion set up in this way fulfils the `conservation of $\mu$'
\eqref{datamodel:conservation-of-mu}.
Note that the factor $\kappa$ can be absorbed into \mixingMatrix\ which we will
assume from now on.
Then, $\mixingMatrix_{ix}$ has the meaning of the rate of events in data bin $i$
per $\e^{b\,s_x}$ in signal space.

It is now in order to summarise in brief our data model.\index{data model}
\begin{itemize}
	\item The log-density signal is a Gaussian field with covariance
	matrix \prop, such that $P(s)=\gauss{s}{\prop}$.
	The matter distribution is modelled by a log-normal \PDF, such that the
	density in $V_i$ is $\rho_i=\rho_0e^s$.
	\item The response is given by $\mu_i=\sum_j \mixingMatrix_{ij}\e^{b\,s_j}$.
	\item The data likelihood given the signal is
	$P(d|s)=\prod_i\frac{\mu_i^{d_i}\,\e^{-\mu_i}}{d_i!}$.
\end{itemize}
It is evident that the complexity of this system strongly depends on the
structure of \mixingMatrix, and in particular if \mixingMatrix\ itself depends
on the signal $s$ to be reconstructed, or not.

The local data model with $\mixingMatrix = \kappa^\tp\openone$ has been solved
in a fully Bayesian framework by \citet{2009PhRvD..80j5005E}.
Implementations thereof in the field of large-scale structure reconstruction
using the MAP method and a Hamiltonian sampler can be found in
\cite{2009MNRAS.400..183K}, \cite{2010MNRAS.403..589K}\footnote{The
authors of this work use a mathematically similar response
$\mu=\kappa\bigl(1+\mathbf{B}(e^s-1)\bigr)$ that could also be used to model
spatial distortion. Unfortunately, the authors do not mention the
interconnection between their general scale-dependent bias $\mathbf{B}$ and a
point spread function.} and \cite{2009arXiv0911.2498J,2009arXiv0911.2496J},
respectively.
There are of course also other survey based reconstructions of the large scale
structure.
Such can be found in
\cite{%
1990ApJ...364..370B,
1991ApJ...372..380Y,
1995ApJ...454...15S,
1996ApJ...461L..17B,
1997MNRAS.287..425W,
1999AJ....118.1146S,
1999MNRAS.303..179N,
2000ApJ...535L...5H,
2001ApJ...550...87G,
2004MNRAS.352..939E,
2004ogci.conf....5V,
2005ASPC..329..135H,
2005MNRAS.356.1168P,
2006MNRAS.373...45E}. 

\subsection{Setting up a reconstruction problem}
In order to test the map making techniques on the log-normal density field with
Poissonian noise, we generate mock data from a known signal and try to
reconstruct the signal.
Aiming at a proof of concept, it is in order to keep the complexity of the
numerical simulation low, hence we restrain ourselves to a 1D case.
This restriction is not fundamental, because the same algorithm can also be
applied to higher dimensional problems.
In a sense the topologies of the spaces involved are encoded in \prop\
and \mixingMatrix, but as we will see below the inner structure of \prop\
and \mixingMatrix\ are a priori irrelevant for our method on an abstract
level\footnote{It may however have an impact on convergence speed, but these
details are left for further investigation.}.
The only downside of multidimensional reconstructions is that the dimension of
the field vector space grows like $N^3$ where $N$ is the number of pixels along
the any direction.
So even moderate resolutions have severe impact on the performance of any
calculation and demand extensive optimisation.
Therefore we contend ourselves with $N$ pixels in one dimension for this
conceptual work.

In order to be free of boundary effects we choose a ring-like topology for our
reconstruction, such that the first and the last pixel are direct neighbours.
This ring-like topology is also not necessary since boundary effects can
be treated naturally, but it simplifies the set-up.

The first step is to choose an adequate \powsp\ for the signal.
Since we are mostly aiming for the principles of reconstructions with
spatially distorted data, the details of the power spectrum are not essential
here as long as most of the power is concentrated on the large scales.
Inspired by the Yukawa potential from quantum electro-dynamics
\citep{1995iqft.book.....P} which mediates a force with a range of $l_c$ or --
in a different language -- introduces correlation with a characteristic length
scale $l_c$ we try $P(k)\propto \frac{1}{k^2+l_c^{-2}}$.
However, in order to have a more structured looking signal, we introduce a boost
term which selectively amplifies power on large scales:
\begin{equation}
	P(k) \propto \frac{0.2+\e^{-0.008\,k}}{k^2+l_c^{-2}}
	\label{mocksig:powerspectrum}
\end{equation}
All reconstructions in this work were done for a correlation length
$l_c=0.05\,L$ where $L$ is the length of the simulation interval.

As the \powsp\ is the diagonal of the Fourier transform of the covariance
matrix, this allows to easily compute
\begin{equation}
	\prop_{xy} = \int dk\,\e^{2\pi\imath(xk-yk)}\widehat{P}(k)
		= \mathcal{F}^{-1}\bigl[\widehat{P}\bigr](x-y)
	\label{mocksig:covmat-from-powspec}
\end{equation}
Here $\mathcal{F}^{-1}\bigl[f\bigr](x)$ must be read as the inverse Fourier
transform of $f$ evaluated at $x$ and $\widehat{P}(k)$ stands for the matrix
with $P(k)$ on the diagonal.
Similarly, we can compute
\begin{align}
	\label{mocksig:inv-covmat} \prop^{-1}_{xy} & =
		\mathcal{F}^{-1}\bigl[1/\widehat{P}\bigr](x-y)\text{, and}\\
	\label{mocksig:sqrt-covmat} \rprop_{xy} & =
		\mathcal{F}^{-1}\bigl[\widehat{P}^{1/2}\bigr](x-y)
\end{align}
which we need for the evaluation of $H_d$ and to generate mock signals,
respectively.
In \appendixname~\ref{appendix:mocksig} we describe how a random signal with
given covariance matrix \prop\ can be constructed from a vector of random
numbers using \rprop, the root of \prop.

\subsubsection{Naïve data inversion}\label{mocksig:naive-map}
As competitor for the fully Bayesian reconstruction, we also construct an
unoptimised filter, namely a non-Bayesian reconstruction, that we call \mnaive.
The naïve data inversion is not as straight-forward as it may seem at first
sight.
Although the performance of \mnaive\ is extremely poor\footnote{Therefore the
subscript `na' for naïve} it is still worthwhile to investigate the problems
of direct data inversion in order to understand why this approach is doomed to
fail.

In principle, naïve data inversion seems sensible, as the data $d$ trace the
response $\mu$ which can directly be inverted to yield $s$.
However, $d$ is not exactly $\mu$ because of the Poissonian sampling of the
response.
As a non-Bayesian analysis neglects the signal covariance for \mnaive, no
information on the true underlying $\mu_i$ can be inferred from adjacent pixels.
Hence there is no way to tell if the data point $d_i$ is lower or higher than
the expected $\mu_i$, and the only choice is to make the approximation $\mu_i
\approx d_i$.
Furthermore, when the distortion depends on the signal, there is no way to
consistently incorporate bin mixing in the data inversion, such that one has to
approximate the distorted response by the local response
\eqref{def:datamodel:mu-local}.
This leads to
\begin{equation}
	\widetilde{m}_{na}(d_i) = \frac{1}{b}\log\Bigl(\frac{d_i}{\kappa_i}\Bigr)
	\label{mocksig:non-bayesian:m-naive}.
\end{equation}
However, this is only possible for pixels with $d_i > 0$, and a generic guess such as
$\widetilde{m}_{na}(d=0) = 0$ for those pixels cannot be correct, as for high
galaxy densities with $w_i\,V_i\,\bar{\rho}_{e}^{(0)}=\kappa_i>1$ the estimated
signal is negative even for $d_i = 1$.
A way out is to allow for Bayesian reasoning in this case and ask for the most
probable field strength $s_i$ given its variance and $d_i=0$\footnote{%
The careful reader may argue, why to distinguish between the two cases $d_i=0$
and $d_i\neq0$.
The same argument as for $d_i=0$ can also be applied to $d_i\neq0$ leading to
the similar result $\widetilde{m}_{na,i}=d_ib\,\prop_{ii}-\frac{1}{b}\plog\left(
\kappa_ib^2\prop_{ii}\,\e^{d_ib^2\prop_{ii}}\right)$.
However, in most situations this result is very close to
\eqref{mocksig:non-bayesian:m-naive}, so we decided to stick to the more
data-oriented non-Bayesian solution.}.
This leads to
\begin{equation}
	\widetilde{m}_{na,i} = \begin{cases}
		\frac{1}{b}\log\Bigl(\frac{d_i}{\kappa_i}\Bigr) & d_i > 0 \\
		-\frac{1}{b}\plog\left(w\,b^2\prop_{ii}\right) & d_i = 0,
	\end{cases}
	\label{mocksig:inverse-data}
\end{equation}
where $\plog(z)$ is the Lambert $\plog$-function (i.e.~the root of $z=w\,\e^w$).
As it stands $\widetilde{m}_{na}$ is however a bad estimator for the signal as
the noise from the Poissonian sampling is very present and the true signal is
known to be smooth.

It is common practice to apply a smoothing procedure to the inverted data.
However, the shape of the smoothing kernel introduces new free parameters and
there is no generic setting for it in non-Bayesian reasoning.
Allowing for the use of the covariance \prop\ gives again a back-door, as it can
be used to smooth the inverted data.
We choose to convolve with \rprop, the root of \prop, as it is more localised
and therefore the convolved map reflects the data better.
Hence the final non-Bayesian map is given by\index{map!naïve}
\begin{equation}
	\mnaive = \rprop\widetilde{m}_{na}
	\label{mocksig:mnaive}
\end{equation}
Even though \mnaive\ gives really poor guesses for the underlying
signal $s$, our tests have shown that smoothing with a top-hat filter performs
even worse.
This also justifies our choice of \rprop\ as smoothing kernel.

We also like to stress, that even this naïve map construction had to rely on
some Bayesian elements in that signal prior information was necessary for
treating the $d_i=0$ case and setting up optimal smoothing.
Besides, at a closer look the smoothing procedure is questionable, because it
introduces a hidden prior\index{prior!hidden} for a signal with a certain
correlation length or \powsp.
But instead of introducing a hidden prior, one should rather be upright,
make clear where the \powsp\ for the signal enters the reasoning,
and include the prior for $s$ right from the start as we propose in the next
section.

\subsubsection{Bayesian map-making}\label{section:general:implementation}
Since a full posterior mean as proposed in \ref{section:optimal-map-making} is
not straight-forward for this problem, we have to rely on approximations for
$\expectationWRT{s}{(s|d)}$.
One possibility would be to sample the posterior via a Monte-Carlo Markov chain
(MCMC) method, but this is computationally very expensive and difficult to
understand analytically, and therefore not suited for a proof of concept at
which we aim.
Instead we use the approximation of $\expectationWRT{s}{(s|d)}$ by the classical
map as proposed in section \ref{section:classical-map}.

For this it is sufficient to minimize the probability Hamiltonian of the problem
defined by \eqref{fundamentals:hamiltonian}.
For our problem defined by the likelihood \eqref{def:datamodel:likelihood} and
the signal prior \eqref{def:datamodel:prior} it is given by
\begin{equation}
	\begin{split}
	H_d[s] & = -\log P(d,\,s) = -\log\bigl(P(d|s)\,P(s)\bigr)\\
	& = 1/2\, \NormWithMetric{s}{\prop^{-1}} -
		\sum_i d_i \log \mu_i + \sum_i \mu_i + \log \bigl(\prod_i d_i!\bigr)\\
	& \cong \frac{1}{2}\NormWithMetric{s}{\prop^{-1}} -
			\sum_i \bigl(d_i \log \mu_i - \mu_i\bigr),
	\end{split}
	\label{conjgrad:H}\raisetag{12pt}
\end{equation}
where we have shifted the Hamiltonian by a constant in the last step.

As the classical map \MAP\ aims at minimizing the Hamiltonian, we can use
efficient multidimensional minimization techniques.
In particular, we use the conjugate gradient method\footnote{For an introduction
to the method of conjugate gradients see \cite{shewchuk}} in the
Broyden-Fletcher-Goldfarb-Shanno variant as provided by the GNU Scientific
Library\footnote{The GNU Scientific Library is available from
\href{http://www.gnu.org/software/gsl}{http://www.gnu.org/software/gsl}} to
minimize $H_d[s]$.

The conjugate gradient method needs the derivative of the function to be
minimized, so we calculate
\begin{equation}
	\variation{H_d}{s} = \prop^{-1}\cdot s - \biggl(\frac{d}{\mu}-1\biggr)^\tp\cdot \variation{\mu}{s}.\label{conjgrad:grad-H}
\end{equation}

In the case where \mixingMatrix\ is independent of $s$ the gradient of $\mu$
is rather easily computed
\begin{equation}
	\variation{\mu_i}{s_k} = \mixingMatrix_{ik}\,b\,\e^{bs_k}.
	\label{conjgrad:grad-mu}
\end{equation}

\subsubsection{Error estimation}\label{section:errorestimation}
The signal reconstruction alone is only of little use, as it contains no
information on the reliability of the reconstruction.
An approximation of the correct error intervals for \MAP\ can be obtained via
approximation of the posterior with a Gaussian.
Therefore, we Taylor-expand the probability Hamiltonian up to second order in
$s$ and obtain $H_d[s]\approx H_0+\frac{1}{2}
\NormWithMetric{(s-\MAP)}{\propD^{-1}}$.
The Gaussian approximation of the posterior is then given by $P(s|d)\approx
\gaussSymbol{s-\MAP}{\propD}$.
With the generating functional from
\eqref{fundamentals:generating-functional-gaussian} it is then easy to
calculate the expected deviation from \MAP
\begin{equation}
	\expectationWRT{\delta s_x\,\delta s_y}{(s|d)} =
	\left.\frac{\delta^2}{\delta J_x\,\delta J_y}\log Z_d\right|_{J=0}
	= \propD_{xy}
	\label{errorestimation:gaussian},
\end{equation}
where $\delta s \equiv s - \MAP$.
One should keep in mind that this procedure gives only an approximation of the
real one-$\sigma$ confidence interval due to the Gaussian approximation of the
posterior and also does not display the cross-correlation between errors at
different locations.
Visual inspection of the error bars obtained for our example figures shows
that they are neither too large nor too small and are therefore a good
approximation of the correct error bars.

In our case where the Hamiltonian is given by \eqref{conjgrad:H} one finds
\begin{equation}
	\propD^{-1}_{mn} = \prop^{-1}_{mn}
	+\sum_i\frac{d_i}{\mu_i^2}\variation{\mu_i}{s_m}\variation{\mu_i}{s_n}-\sum_i\Bigl(\frac{d}{\mu}-1\Bigr)_i\frac{\delta^2\mu_i}{\delta
	s_m \delta s_n}
	\label{errorestimation:my-D}
\end{equation}
where we have not implied any summation for clarity.

If \mixingMatrix\ depends on the signal we have to make the simplifying
assumption that it varies only slowly with $s$ such that its contribution to the
gradient of $\mu$ can be neglected.
The reasons for this simplification is mainly that for our specific
$\mixingMatrix[s]$-model developed in \ref{section:ld-distortion-model} this
derivative is computationally extremely expensive to calculate, let alone the
second derivative of \mixingMatrix.
Therefore, one gets
\begin{align*}
	\variation{\mu_i}{s_m} & = b\,\mixingMatrix_{im}\e^{b\,s_m} &
	\frac{\delta^2\mu_i}{\delta s_m \delta s_n} & =
	\delta_{mn}\,b^2\mixingMatrix_{im}\e^{b\,s_m},
\end{align*}
which can be used in \eqref{errorestimation:my-D}.

\subsection{Photometric redshifts}\label{section:photometric-redshift}
We apply our reconstruction algorithm to different cases, one of which is the
reconstruction of dark matter density from galaxy counts whose redshift is
determined photometrically.
Photometric redshift measurement is based on the apparent colours $c$ of
galaxies rather than their full spectra, and schemes to assign a redshift $z$
depending on the measured $c$.
However, as \citet{2000ApJ...536..571B} points out, there is no unambiguous
colour to redshift mapping even if more and more band filters are used.
He argues that instead of a strict mapping $z=z(c)$, one has to work with the
full posterior $P(z|c)$ that a galaxy with colour $c$ actually has redshift
$z$.

\citet{2009ApJ...700L.174W} treats this problem by drawing a random $z_{MC}$
from $P(z|c)$ for each galaxy and continues his calculation with this $z_{MC}$.
However, he is aware that this procedure only works if the number of galaxies
per colour bin is large.
So we choose a density field reconstruction from photometric redshift data as a
testing ground for our method.

\subsubsection{Distortion matrix for photometric redshifts}%
	\label{section:photometric-distortion}
The distortion matrix for photometric redshift distortions maps from redshift
into colour space and must be set up according to $\mixingMatrix_{cz} = P(c|z)$.
\begin{figure}
	\includegraphics{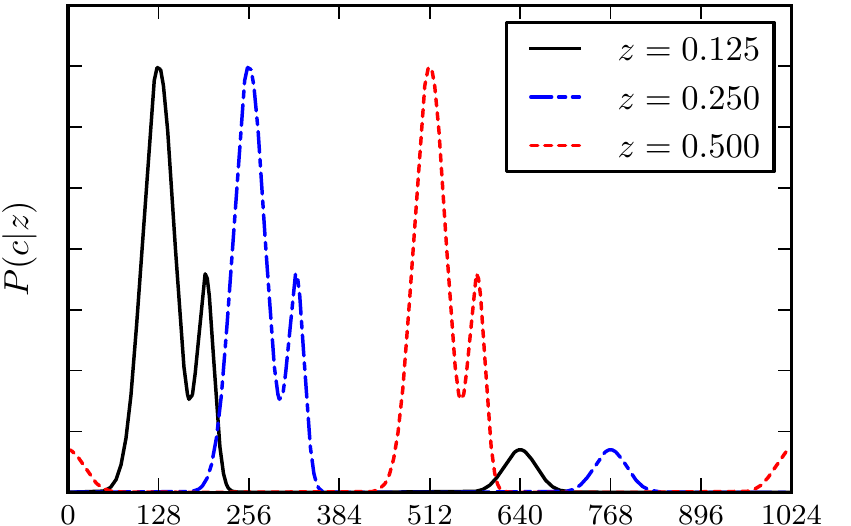}
	\caption[Colour likelihood for a given redshift]%
		{The probability distribution of observed colour $c$ for three different
		redshifts.
		The x-axis is the number of the colour bin, the galaxy may be put into.
		Here wee use $1024$ colour and redshift bins and the maximal redshift
		bin corresponds to $z=1$.
		Note that the shape of the distribution is assumed to be independent of
		$z$ and is only shifted towards higher colour values.
		This is a simplification \wrt\ real photometric redshift probabilities,
		where the shape of the \PDF\ \emph{does} depend on $z$.
		As we use a cyclic topology for this test-case, $z=1$ corresponds to
		$z=0$, and $c=0$ to $c=1024$, respectively.
		Therefore, the catastrophic outliers for $z=0.500$ (dotted red) reappear
		at the low colour values on the left.}
	\label{fig:photometric-prob}
\end{figure}
The features we want to test for are asymmetric shape of \mixingMatrix\ and the
robustness to \emph{catastrophic outliers}\index{catastrophic outliers}.
As this is intended as a test-case, we make some simplifying assumptions about
$P(c|z)$.
First, we assume that the shape of $P(c|z)$ remains fixed when going to higher
redshift and second we neglect the effect of the spectral type on the colour
\PDF.
Catastrophic outliers are modelled by a small Gaussian \PDF\ contribution which
is offset from the main peak by half of the simulated interval length and whose
height was chosen to be one tenth of the main peak.
It should be stressed however, that once \mixingMatrix\ is set up in a
realistic way, no change is needed for the algorithm.
In \figref{fig:photometric-prob} we show the $P(c|z)$ we use and it should be
explicitly said that this is \emph{not} a real-world $P(c|z)$ but one that we
designed to just look similar to a real-world $P(c|z)$ as in
\cite{2000ApJ...536..571B}.

\subsubsection{Reconstruction of redshift space matter distribution from colour
	space data}\label{section:ph:results}
For our reconstruction test we set up our simulation on an interval of length
$L=1$ split into 1024 evenly sized pixels.
As we are not interested in boundary effects here, we set the window function to
unity for the whole interval.
\begin{figure}
	\includegraphics{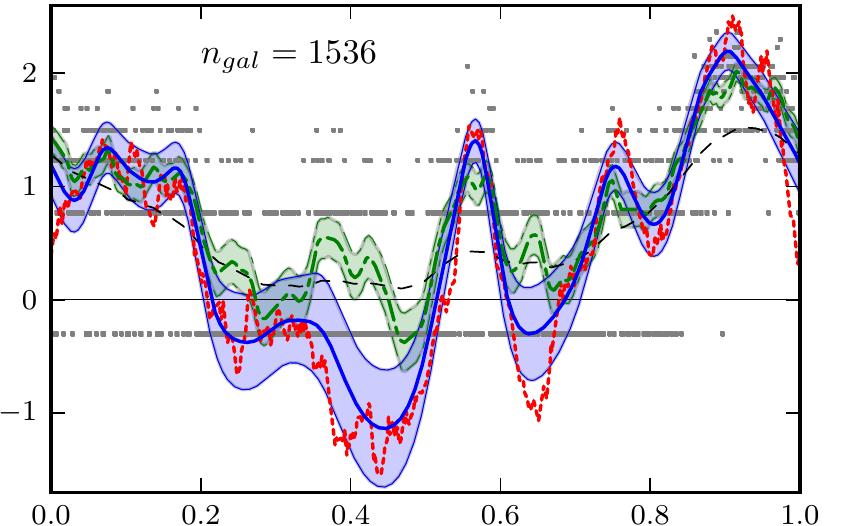}
	\caption[Reconstruction of a signal from photometric redshift data]%
		{Reconstruction for a bias $b=1.5$ and $\bar{\rho}_{gal}=1000$ of a
		signal (red dotted curve) from data (grey dots) with Poissonian noise
		and typical spatial distortions as they occur in photometric redshift
		measurement.
		The data is where the point-wise inversion from
		\eqref{mocksig:inverse-data} would place it and the thin evenly dashed
		black line shows the naive map \mnaive.
		The smooth blue line shows the classical reconstruction with the correct
		distortion matrix, the dash-dotted green line shows \MAPu, a MAP map
		without the distortion of colour space taken into account.
		$1\sigma$ error levels are indicated by thin lines.}
	\label{fig:photometric-big}
\end{figure}
\begin{figure*}
	\includegraphics{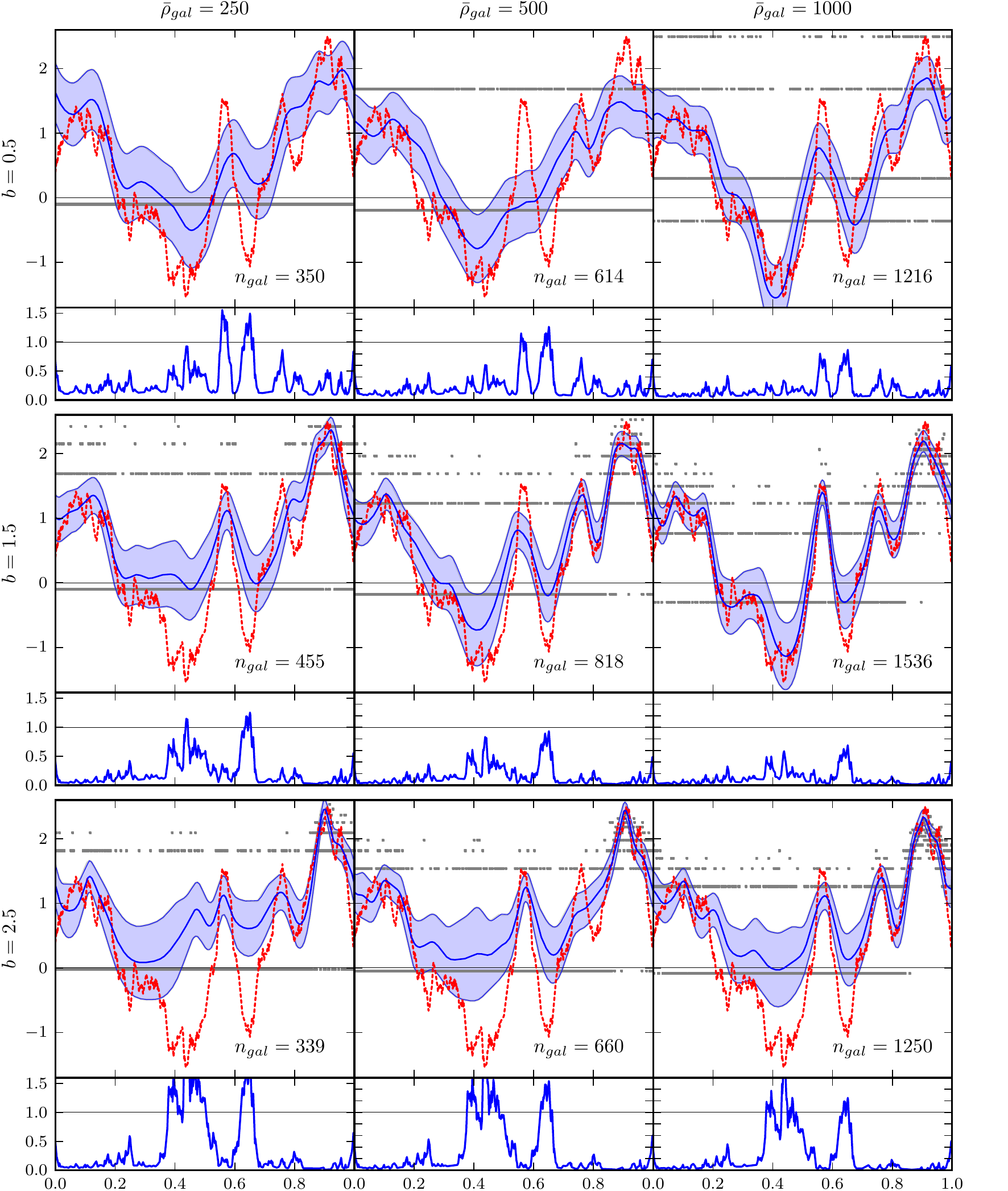}
	\caption[Table of nine reconstructions from photometric redshift data]%
	{The top part of the nine panels show reconstructions of the same
	signal (dotted red) from data (grey dots) with Poissonian noise and typical
	spatial distortions as they occur in photometric redshift measurement.
	The data is where the point-wise inversion from \eqref{mocksig:inverse-data}
	would place it.
	The blue line shows the classical signal reconstruction as proposed by our
	method with its $1\sigma$ error contours (thin blue lines).
	The bottom part of the nine panels show the characteristic residuals as
	defined by \eqref{def:characteristic-residual} for $500$ different data
	realisations for this signal.}
	\label{fig:photometric-table}
\end{figure*}

In \figurename s \ref{fig:photometric-big} and \ref{fig:photometric-table} we
show reconstructions of the signals with unit variance
$\expectationWRT{s_x^2}{(s)}=1$ from photometrically distorted data.
In \figref{fig:photometric-table} we also show the \emph{characteristic
residuals}\index{characteristic residual} for this single signal \strue\ when
reconstructed by this technique, defined as
\begin{equation}
	\res^{(\strue)} = \expectationWRT{(m_d-\strue)^2}{(d|\strue)}
	\label{def:characteristic-residual}
\end{equation}
where $m_d$ is the map as it is reconstructed from the data $d$.
In practice, we do this average for 500 different data realisations of the same
signal \strue.
This shows in which regions the algorithm generally performs badly for a
specific signal and where it does well independently from the data realisation.
One may notice in \figref{fig:photometric-table}, that in actual data
realisations the galaxy numbers vary for different biasses while keeping
$\bar{\rho}_{gal}$ constant.
This has to be expected because the average galaxy density is only the same for
different biasses, when averaged over all signal realisations.
This may affect the comparability of the characteristic residuals for different
biasses, but not for the same bias and different galaxy densities.

The first thing to notice about the reconstructions is that in all cases they
are pretty smooth even though the noise in the data is considerable and also the
original signal has far more power on small scales.
This is against common knowledge that MAP maps pick up a lot of noise.
What prevents this in our case is the distortion matrix which act as a smoothing
operation on the observed data such that \MAP\ itself becomes smooth again.

In \figref{fig:photometric-big} we also show a classical reconstruction \MAPu\
that does not take the distortion of data space into account, i.e.~it assumes
that $\mixingMatrix = \kappa^\tp\openone$.
Compared to our full reconstruction \MAP\ one can see that neglecting the
colour space distortion results in severe difficulties to get even the shape
right.
Instead of voids \MAPu\ even sees peaks at $0.4$ and $0.65$.
Although large peaks are detected, their height is slightly underestimated
because some events are scattered away which \MAPu\ is not aware of.
The full reconstruction on the other hand can even reshape the void region
located at $0.4$, despite the abundance of events in that region originating
from the large peak at $0.9$.
Looking at the characteristic residuals in \figref{fig:photometric-table}
reveals that this is true for most data realisations.

In \figref{fig:photometric-table} we show reconstructions of the same signal for
different simulation parameters and their average residuals.
In the reconstructions, one can make out two obvious trends for our simplified
model:
\begin{itemize}
	\item The higher the galaxy density, the better the overall reconstruction
	(note in particular the decreasing level of $\res^{(\strue)}$).
	\item The higher the bias the better becomes the reconstruction of peaks,
	but the worse becomes the reconstruction of void regions.
\end{itemize}
Both observations are not surprising, since higher galaxy density means that the
response is sampled with less Poissonian noise, so one expects the
reconstruction to become better.
Higher bias on the other hand sharpens the contrast such that peaks are
selectively sampled with high accuracy whereas the galaxy density in void
regions is reduced.
This trend is also reflected by the error bars which tighten up in overdense
regions for biasses larger than one, but not so for $b=0.5$.

The example for $b=0.5$ and $\bar{\rho}_{gal} = 500$ shows that data variance
can make a big difference.
Although the number of galaxies is nearly twice as large as in the panel to its
left, the peak at $0.55$ is hardly detected at all.
Looking at the residuals reveals that this is simply an artefact of a lucky data
realisation for the low density case and an unlucky one for the middle density
case.

Looking at the characteristic residuals one notices, that for $b=0.5$ the
general trend of the signal (i.e.~the largest Fourier modes) can be detected
even for very low galaxy densities, while sharp peaks and ditches are even
difficult to resolve for $\bar{\rho}_{gal}=1000$.
For $b=1.5$ and $b=2.5$ the overdense regions can be resolved even for low
density of events, while the voids are difficult to reshape for low galaxy
densities.

It should be mentioned however, that the poor resolution of voids is also a
problem of this specific example, because every void region has an overdense
region that scatters events into the void ($0.9$ into $0.4$ and $0.15$ into
$0.65$).
Had we chosen an example where two void regions scatter into each other, the
resolution of these voids would be much better.

\begin{table}
\centering
	\caption{%
		As an indicator for the quality of different reconstructions this table
		lists $\expectationWRT{\Lpnorm{m-\strue}{2}^2}{(\strue)}$, the average
		$L_2$-distance from reconstruction $m$ to true
		signal $\strue$, where the $s$-average runs over 500 random signal
		configurations.}
	\begin{tabular}{c|c||c|c|c}
		&& $\bar{\rho}_{gal} = 250$ & $\bar{\rho}_{gal} = 500$ & $\bar{\rho}_{gal} = 1000$ \\
		\hline
			& \mnaive &1.03 &1.03 & 0.78 \\
		$b=0.5$	& \MAPu &0.34 &0.27 & 0.23 \\
			& \MAP &0.31 & 0.15 & 0.16 \\
		\hline
			& \mnaive &0.68 &0.63 &0.58 \\
		$b=1.5$	& \MAPu & 0.36 & 0.33 & 0.32 \\
			& \MAP &0.25 & 0.19 & 0.15 \\
		\hline
			&\mnaive &0.74 &0.70 & 0.68 \\
		$b=2.5$	& \MAPu & 0.60 & 0.57 & 0.54 \\
			& \MAP &0.41 & 0.34 & 0.28
	\end{tabular}
	\label{tab:photometric:avg-residuals}
\end{table}
In \tableref{tab:photometric:avg-residuals} we list the average residuals
of 500 reconstructions of different signals.
The naive map \mnaive\ is in all cases the worst reconstruction.
Especially for low galaxy densities it is painfully close to the average
residual of a zero map, which is expected to be unity for a Gaussian signal with
unit variance.
This tells above all one thing: for a reconstruction in the low signal to noise
regime one must include some knowledge about the signal.
For a low bias \MAPu\ and \MAP\ have roughly the same error level when the
galaxy density is low.
When $\bar{\rho}_{gal}$ rises, both maps become better but what surprises is
that \MAP\ seems to saturate over $\bar{\rho}_{gal}\ge 500$ at an error level of
$\approx 0.15$.
Considering the fact, that the distortion of the signal inevitably destroys some
information, the question arises, if this is already the optimum.
Therefore, we let the same signals be processed without noise in the
response, i.e.~with perfect data and found for $b=0.5$ that without Poissonian
noise \MAP\ has an average residual of $0.09$ and \mnaive\ of $0.27$.
So we see that for the signals with $\bar{\rho}_{gal}=500$ and
$\bar{\rho}_{gal}=1000$ there is still is some margin to be gained by even
higher galaxy densities, but not much.
It is remarkable, that \mnaive\ with noiseless data comes not even close to the
performance of \MAP\ with medium galaxy density.

There is one anomaly worth noticing, namely the increase of the residuals
from $b=1.5$ to $b=2.5$ observed for all maps but \MAPu\ in particular, which
comes from catastrophic outliers scattering into void regions.
Since count rates for $b=2.5$ are only large at big overdensities and remain
small even for smaller peaks, \MAPu\ makes a small peak from scattered events
in voids.
The same holds also for \mnaive, but this map does not react as quickly on few
events as \MAPu\ does, which explains the modest deterioration of \mnaive\ and
the striking difference for \MAPu.

\subsection[Reconstruction from multiple data sets]%
	{Signal inference from independent data sets}%
	\label{section:combined-data-sets}
In real world surveys like SDSS there might be both available, accurate
spectroscopic galaxy redshift and abundant less certain photometric redshift
data.
Can the former help to better localise the latter?

Our formalism allows to combine both data sets by defining
\begin{equation}
	d = \left(d^{(1)},\,d^{(2)}\right) =
			\left(\mixingMatrix^{(1)},\,\mixingMatrix^{(2)}\right)^\tp\kappa\,\e^{b\,s}
	\label{def:combined-data-vector}
\end{equation}
and business is as usual.

\begin{figure}
	\includegraphics{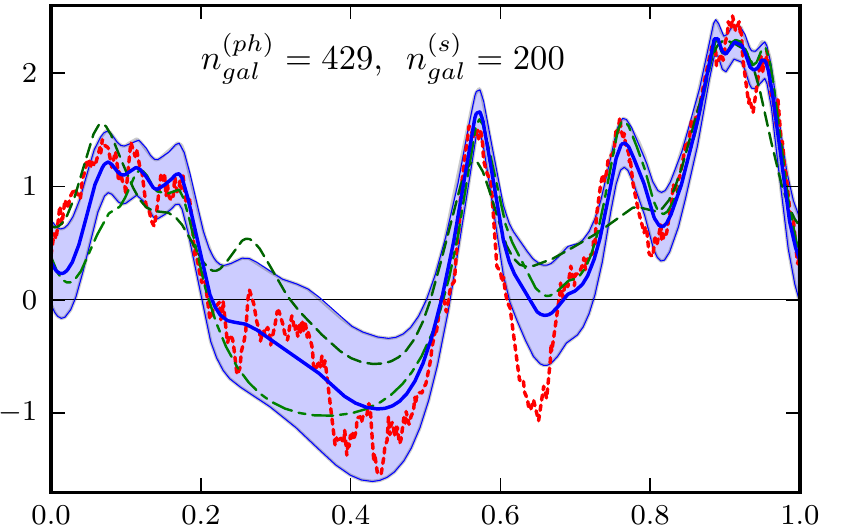}%
	\caption[Reconstruction from two independent data sets]%
	{Classical reconstruction of a signal (red dotted line) from two
	independent data sets (data points not shown) for a bias of $b=1.5$,
	$\bar{\rho}^{\,ph}_{gal} = 250$ and $\bar{\rho}^{s}_{gal} = 60$.
	The reconstruction from photometric data alone is shown as a dark-green
	evenly dashed curve, the reconstruction from spectroscopic data as a green
	dash-dotted curve.
	The combined analysis of both data sets is shown as an blue smooth curve.
	The $1\sigma$ error level of the combined reconstruction is indicated by the
	thin blue solid line.
	For the other reconstructions it was omitted for clarity.}%
	\label{fig:cdata:bigmap0098}
\end{figure}
In principle this allows to combine any two data sets, but now consider the case
where $d^{(1)}$ are data from photometric redshift measurement and $d^{(2)}$
from spectroscopic redshift measurement.
As such, we model $\mixingMatrix^{(1)}$ as in section
\ref{section:photometric-distortion}.
Since spectroscopic redshift measurement is far more accurate than photometric
redshift measurement, we assume spectroscopic redshift measurements to be exact,
i.e.~$\mixingMatrix^{(2)}=\openone \kappa$ with $\kappa$ being the zero-response
\eqref{def:datamodel:mu-local}.

In \figref{fig:cdata:bigmap0098} we show a reconstruction for one data
realisation $d^{(1)}$ and $d^{(2)}$ with a bias $b=1.5$.
The combined reconstruction $\MAP^{(s+ph)}$ mostly follows the reconstruction
from spectroscopic data $\MAP^{(s)}$, especially in overdense regions such as
from $0.8$ to $1.0$ and $0.5$ to $0.6$.
This is an expected behaviour, since in this case $b=1.5$ and therefore
overdense regions are accurately sampled even with the low average galaxy
density $\bar{\rho}^{s}_{gal} = 60$.
And since the spectroscopic data have no spatial error, they dominate the
reconstruction in this regime.
However, in regions where spectroscopic galaxies are rare, the combined
reconstruction sometimes deviates substantially from $\MAP^{(s)}$ (most
conspicuous from $0.2$ to $0.4$).
The crucial point to note is that the combined reconstruction is not an average
of the two reconstructions $\MAP^{(s)}$ and $\smash{\MAP^{(ph)}}$ as can be seen
at $0.4$ to $0.5$ and $0.6$ to $0.7$ where the combined reconstruction lies
below the others.
Therefore, the optimal combination of the two datasets is non-trivial and
non-linear.

The statistics for 500 different signal realisations (not shown) confirm that
the combination is more than just a superposition of the two single-dataset
reconstructions.

\subsection{X-ray astronomy via coded mask telescopes}\label{section:coded-mask}
Thanks to the generic structure of our approach, we can also apply it to a
completely different problem.
One such problem is the detection of extended sources in coded mask aperture
systems.

Coded aperture systems were originally proposed by \cite{1968ApJ...153L.101D}
for the purpose of detecting point-like X-ray sources.
In this scheme an absorbing plate with a pattern of pinholes is placed in front
of a detector and the shadow of this plate on the detector allows with the
knowledge of the plate pattern to unfold the count distribution and infer the
positions of the X-ray sources.
However, this technique becomes much more difficult, when the light sources are
extended and not just point sources.

We now demonstrate that it is in principle possible to map out extended sources
with our method, when count rate and bias are high enough.
Adapting our method to this problem, the mixing matrix in this case must be the
coded mask pattern that lets light pass for open pixels and shields it
completely for dark pixels.
Hence we only need the pattern of a coded mask for our distortion matrix.
Today's coded masks have an optimised pattern, but for our purpose the
originally proposed random pattern \citep{1968ApJ...153L.101D} of blocking and
transparent pixels is sufficient.

\begin{figure}
	\includegraphics{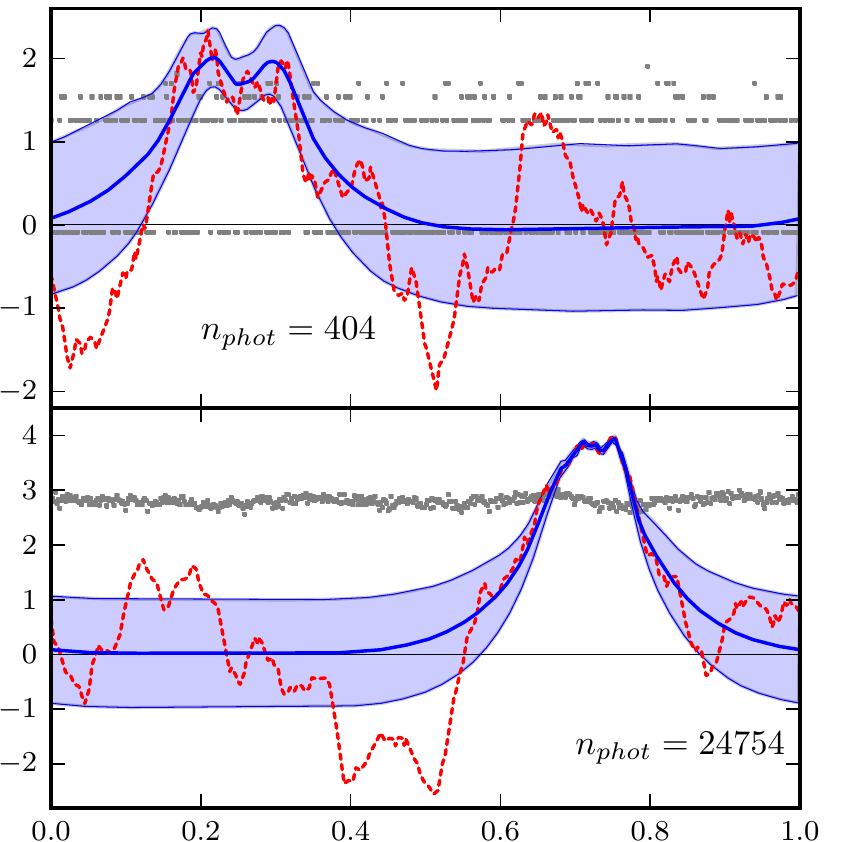}
	\caption[Two reconstructions from coded mask data]%
	{Reconstruction of a signal (red dotted line) from data (grey dots)
	that mimic the behaviour of coded mask apertures with a bias $b=2.5$ and an
	average photon count rate $\bar{n}_{phot}=500$.
	The \MAP\ reconstruction (blue smooth curve) shows only the largest peak of
	the signal, and some of its substructure.
	$1\sigma$-error levels are indicated by thin lines.
	Being completely useless, the naïve map \mnaive\ was omitted.}
	\label{fig:cm-big}
\end{figure}
In \figref{fig:cm-big} we show two reconstructions on an interval with 512
pixels from data obtained via a random pattern coded mask.
In this set-up it is only possible to detect the largest peak, therefore we use
a rather large bias of $b=2.5$.
It is remarkable that a surprisingly low number of photons $n_{phot}=404$ is
sufficient to detect the peak at position $0.25$ in the top panel and even some
bits of its substructure.
In regions where the algorithm can not see the signal, the error bar widens up
to the interval +1 to -1.
This is a good consistency check, because this is expected for a signal with
unit variance when no information on the signal is available.

The lower panel in \figref{fig:cm-big} shows a reconstruction from a signal with
an extremely high peak.
Although such high peaks and count rates are quite rare, it is interesting
how much details of the peak can be reconstructed with extremely small error
range.
Note that the smaller peak at $0.15$ in this example -- although comparable in
size to the peak in the top panel -- is not detected at all.
This is due to the overwhelmingly larger brightness of the largest peak at
$0.75$ whose photons hit the same detector, but with an approximately $\e^{b\cdot
4}/\e^{b\cdot 2}\approx 148$ higher rate.
The Poissonian noise from the larger peak overlays with the photons from the
smaller peak, rendering it impossible to detect.

This is however not yet a fully realistic set-up and further effort must be put
into refining this technique, as sometimes false peaks (with narrow error bars)
turn up in the reconstruction.
It is possible that this problem  has already been amended by the invention of
especially designed coded masks which we did not consider here.

\subsection{Reconstruction with $s$-dependent distortion}%
	\label{section:s-dependent-distortion}
We now address the problem where the distortion matrix depends on the signal to
be reconstructed.
The paradigm for this is the measurement of distance via redshift.
Due to the peculiar velocities of galaxies with respect to the Hubble frame
the comoving distance of galaxies can not exactly be calculated from their
redshift.
Note that there needs not be a coordinate transformation from redshift into
real space, because objects with different positions may have the same redshift.
Therefore, the mapping from real to redshift space is not injective, thus not
invertible.

In particular, the gravitational pull of matter overdensities affects the
peculiar velocities of galaxies.
So if one wants to unfold the large-scale matter distribution in the universe by
measuring angular positions and redshifts of galaxies, one has to take the
distorting effects of matter on the redshift space into account.
What makes this problem so particularly demanding is that the distortion
operator, which transforms the real-space matter distribution into the observed
redshift space matter distribution, depends on the real-space matter
distribution itself which is to be reconstructed.

One has to be aware that the assumption that the matter field is sampled by a
Poissonian process will fail at some point when going to ever smaller scales.
Here we show up problems one has to face even in an idealistic set-up where
the forward transformation from real to redshift space is perfectly known.

\subsubsection{A statistical model for redshift space
	distortions}\label{section:ld-distortion-model}
We now address the forward problem of constructing the redshift space galaxy
distribution from a known real space matter distribution and try to keep our
model as simple as possible.
In the following, ``redshift space'' will always refer to our model
redshift space which we construct with the same characteristic features as the
redshift space found in nature.
To that we develop a simple model which bases on a few generic assumptions.
We first distinguish between two regimes: the linear and the non-linear regime
of large-scale structure formation.

The linear regime is dominated by the peculiar velocities from the bulk motion
of matter in the direction of large-scale overdensities.
The matter flow in this regime is described quantitatively by linear
perturbation theory as
\begin{equation}
	\vec{v}_{l} = \frac{2f(\Omega)}{3aH\Omega}\vec{\nabla} \Phi
	\label{zdist:dv}
\end{equation}
where $f(\Omega) \equiv \frac{d\log D_+(a)}{d\log a} \approx \Omega^{0.6}$
and $D_+$ is the linear growth factor of matter perturbations during the matter
dominated era of the universe.
The potential $\Phi$ is determined by
\begin{equation}
	\nabla^2\Phi = 4\pi G\bar{\rho}a^2\biggl(\frac{\rho-\bar{\rho}}{\bar{\rho}}\biggr)
	\label{zdist:dphi}
\end{equation}
where $\bar{\rho}$ is the mean matter density and $G$ is the gravitational
constant \citep[chapter 6]{2005pfc..book.....M}.
The first to point out the connection between linear perturbation theory of
matter perturbations and redshift space distortions was
\cite{1987MNRAS.227....1K}, for an extensive review on the topic we refer to
\cite{1998ASSL..231..185H}.
The bottom line is that linear distortions lead to an amplification of
overdensities and a depletion of underdensities in redshift space.

However, this relation can only be valid on large scales, because small scale
inhomogeneities rather collapse and form a gravitationally bound object for
which linear perturbation theory fails.
Experimentally this manifests in the deviation of the redshift space matter
\powsp\ from the expected \powsp\ of linear perturbation theory
for wave numbers $k \ga 0.15\,h\,\mpc^{-1}$
(e.g.~\cite{2007ApJ...657..645P,2003MNRAS.341.1311S}).
Therefore, when we calculate $v$ from \eqref{zdist:dv}, we first apply a
tophat lowpass filter on the potential $\Phi$ and continue with
\eqref{zdist:dv}.
This way, only the largest modes of the linear velocity field are resolved which
we use in our set-up of the distortion matrix \mixingMatrix.

The non-linear regime sets in when the perturbations in the matter distribution
collapse and form gravitationally bound objects of numerous galaxies.
In redshift space this presents itself as elongated structures along the line of
sight which are called `\emph{fingers-of-god}'\index{finger of god}.
As the gravitational force of any galaxy on any other in the superstructure are
relevant, this is effectively an N-body problem which is known to behave
chaotically.
So in contrast to the linear contribution to the redshift space distortions, the
non-linear contribution to the peculiar velocity cannot be explicitly given.
It is therefore in order, to resort to a statistical approach here, and we
assume that objects in the non-linear regime are completely virialised and that
their velocities have a Boltzmann \PDF.

From the virial law of classical mechanics \citep{1966mech.book.....L} we can
obtain a relation between the gravitational potential and the time averaged
velocity: $\expectationWRT{\vec{v}^2}{t} = -\expectationWRT{\Phi}{t}$.
Assuming $\expectationWRT{\Phi}{t}\approx\Phi$ yields a relation between the
potential and the velocity dispersion.
We now assume that the velocity \PDF\ for the objects in a virialised system is
given by a Boltzmann factor
\begin{equation}
	P(\vec{v}_{nl})\,d\vec{v} \propto \exp\biggl(-\frac{\vec{v_{nl}}^2}{\Phi}\biggr)d\vec{v}.
	\label{zdist:Boltzmann-raw}
\end{equation}
which guarantees the right velocity dispersion.
That non-linear velocity distortions are given by a Maxwellian distribution is
a common assumption used by many authors \citep[e.g.][]{1994MNRAS.267.1020P,%
1995MNRAS.275..483H,1999MNRAS.305..527T,1996MNRAS.282.1381T}.
The other school of thought models the non-linear velocity distribution as
an exponential pairwise \PDF, i.e.~$P(v)=\bigl(2^{1/2}\sigma\bigr)^{-1}
\exp\bigl(-2^{1/2}|v|/\sigma\bigr)$ \citep[e.g.][]{1997ApJ...475..414B,%
1995clun.conf..143H,1998MNRAS.296..191R}.

We now have to blend the two regimes into one general expression.
Therefore, we extend \eqref{zdist:Boltzmann-raw} and make the ansatz
\begin{equation}
	P(v_\|) \propto \exp\biggl(-\frac{(v_\| - v_{l,\|})^2}{2F(\Phi)}\biggr)
	\label{zdist:Boltzmann-final}
\end{equation}
where $F$ is a continuous function to be determined and $v_{l,\|}$ is the
component along the line of sight of the linear velocity field determined by
\eqref{zdist:dv}.
The function $F$ must have the following limiting behaviour in order to
interpolate between the linear and non-linear regime:
\begin{align*}
	\lim_{\Phi\rightarrow-\infty} F(\Phi) & = |\Phi|/2 &
	\lim_{\Phi\rightarrow+\infty} F(\Phi) & = 0.
\end{align*}
Since $F$ stands in the denominator, it must never become exactly $0$, hence it
must also not change sign.
Apart from that, the Boltzmann factor should play little to no role in the
linear regime where $\Phi$ is above a threshold $\Phi_0$ over which we do not
assume objects to be virialised.
In principle, we should therefore require that $F \approx 0$ in these regions,
i.e.~exact measurement.
In practice on the other hand, every measurement comes with uncertainties and
even in the linear regime galaxies can have unpredictable small peculiar
velocities.
Subsuming this as `measurement uncertainties', it makes sense to
require that $F(\Phi) \geq \sigma_0$ which includes this instrument noise
naturally in our formalism.
In other words, we turn the shortcoming of our method to model exact
measurement into the feature to always include a minimal error with variance
$\sigma_0$.
For the non-linear regime we assume that galaxies are only virialised with their
potential excess $\Delta \Phi = \Phi - \Phi_0$, i.e.~$F(\Delta\Phi) =
-\Delta\Phi/2$.

One possibility to smoothen the transition from linearly to non-linearly
dominated regions is by a tilted hyperbola such as
\begin{equation}
	F(\Delta\Phi) = \frac{1}{4}\biggl(\sqrt{\bigl(\Delta\Phi+2\sigma_0^2\bigr)^2 + \tau^2} - \Delta\Phi\biggr) + \frac{1}{2}\sigma_0^2,
	\label{zdist:transition-func}
\end{equation}
which has the advantage of only introducing one more free parameter $\tau$ which
controls the smoothness of the transition.

This allows to write the distortion matrix for the transformation from real to
redshift space as
\begin{equation}
	\mixingMatrix_{ij} \equiv P(z_i|x_j,\,\rho) \propto \exp\biggl(-\frac{(z_i -
				v_{l\|,j})^2}{2F(\Phi_j-\Phi_0)}\biggr).
	\label{zdist:distortion-matrix}
\end{equation}
Altogether we now have five free parameters to control the behaviour of our
self-built velocity dispersion model, namely
\begin{itemize}
	\item the strength of the linear velocity distortions, which is given for a
	cosmological model
	\item the width of the tophat lowpass filter for the linear velocity
	field, which can be determined by analysing the redshift space matter power
	spectrum as $k \la 0.15\,h\,\mpc^{-1}$
	\cite{2007ApJ...657..645P,2003MNRAS.341.1311S}
	\item $\tau$ adjusting the smoothness of the transition from linear to
	non-linear behaviour
	\item $\Phi_0$ for the zero-level of the gravitational potential under which
	non-linear effects kick in
	\item and $\sigma_0$ for any further measurement error.
\end{itemize}
\begin{figure}
	\begin{center}%
		\includegraphics{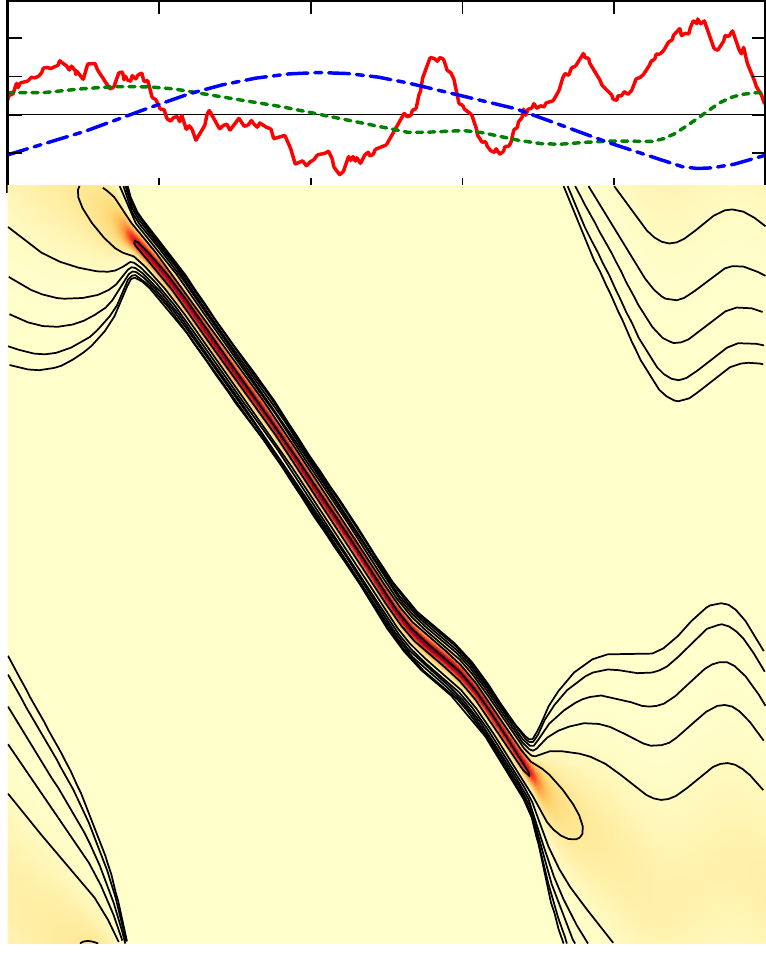}
	\end{center}
	\caption[Distortion matrix for transformation from real to redshift space]%
	{Example for a possible distortion matrix.
	Note that the maximum of the distribution is not necessarily on the diagonal
	due to the linear redshift distortions.
	The log-spaced contours indicate levels of equal height.
	On top the signal (solid red line) that generates this distortion matrix,
	the potential (blue dash-dotted line) and the streaming velocity resulting
	from linear theory (green dotted line) -- all scaled to comparable
	amplitudes.
	Note that the strength of the linear displacement is slightly larger than in
	our simulation to make its effect better visible to the eye.}
	\label{fig:zdist:distortionmatrix}
\end{figure}
As we are aiming at a proof of concept, we also take the gravitational constant,
which adjusts the strength of the non-linear effects, as a free parameter to
generate visible effects with our distortion model\footnote{For completeness we
give the numerical values of our settings: $4\pi Ga^2=0.25$,
$\frac{2\,f(\Omega)}{3aH\Omega}=1.28$, $\tau=5\cdot 10^{-4}$,
$\sigma_0=1.75\cdot 10^{-4}$, $\Phi_0 = 0$ and the tophat filter lets the lowest
1\% of $k$-modes pass.}.
In \figref{fig:zdist:distortionmatrix} we show the mixing matrix as it was used
in the reconstructions of the following section.

As before, we use the conjugate gradient method for minimization of the
probability Hamiltonian.
This method requires the derivative of the Hamiltonian and therefore also the
derivative of $\mixingMatrix[s]$ \wrt\ $s$.
However, for our model a direct calculation of the gradient of \mixingMatrix\ is
not feasible, as it would demand the evaluation of $N_{pix}^3$ difficult to
compute entries of \mixingMatrix, which would pose strong limitations on the
performance, as the evaluation of \mixingMatrix\ already is the bottleneck of
our algorithm.
Therefore we make the approximation, that the contribution of
$\variation{\mixingMatrix}{s}$ is small compared to the other terms in the
gradient of $H$, i.e.
\begin{equation}
	\begin{split}
		\variation{H}{s_k} & = \bigl(\prop^{-1}s\bigr)_k - \biggl(\frac{d}{\mu}-1\biggr)^\tp \variation{\mu}{s_k} \\
		& \approx\sum_j\prop^{-1}_{kj} s_j - \sum_i\biggl(\frac{d}{\mu}-1\biggr)_i
		\mixingMatrix_{ik} \, b\,\e^{b\,s_k}.
	\end{split}
	\label{zdist:R-approximation}
\end{equation}
Although this is an approximation, a simulated annealing process started from
the resulting map never finds a better minimum than this.
Therefore we can safely regard the minimum obtained with the approximated
gradient as the true minimum of $H$.

Similarly, we are forced to make the same approximation when we approximate the
error bars as already mentioned in \ref{section:errorestimation}.

\subsubsection{Results from the reconstruction of the real space matter
	distribution from redshift space data}
\begin{figure*}
	\includegraphics{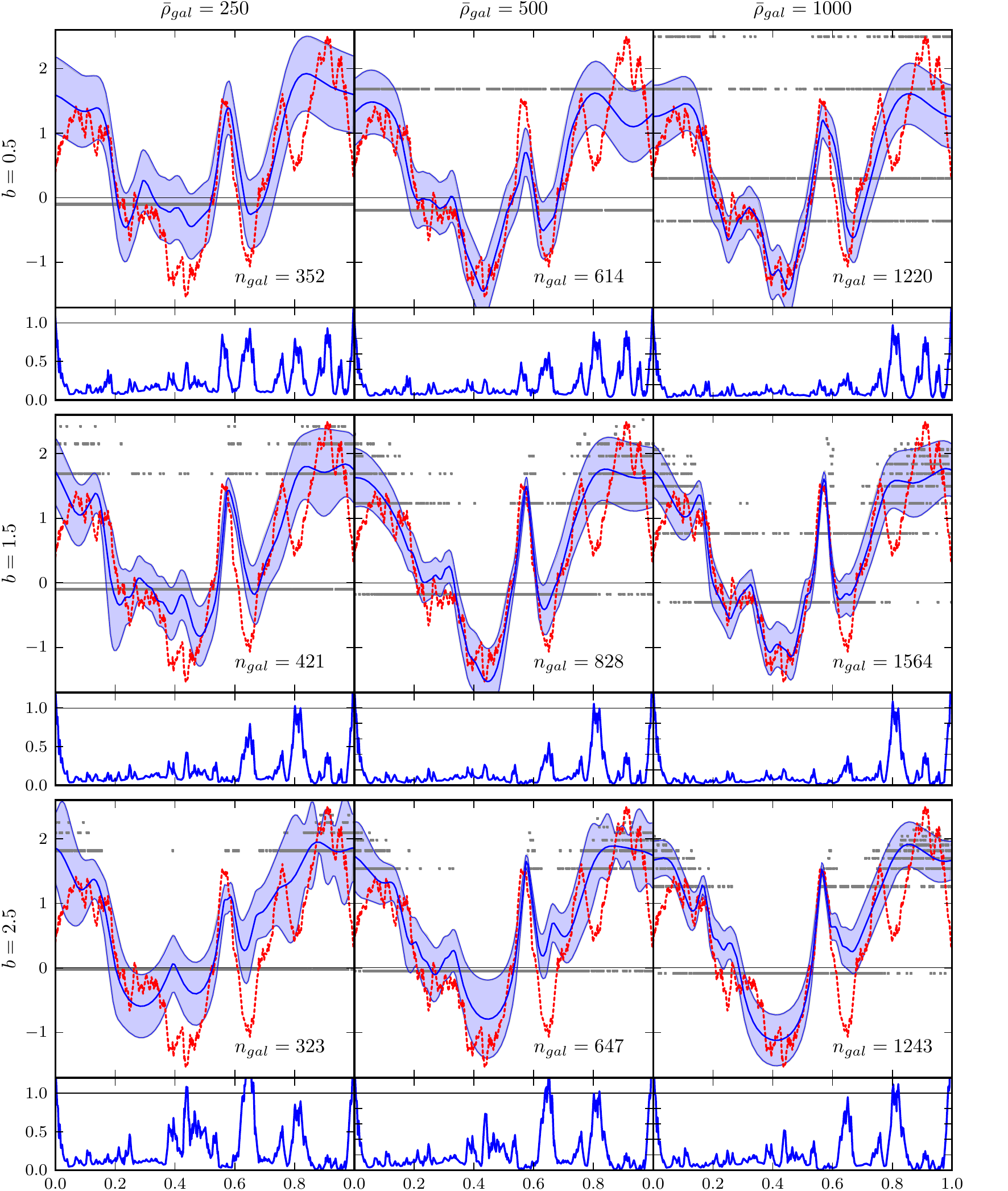}
	\caption[Table of nine reconstructions from redshift distorted data]%
	{The top part of the nine panels show the real space reconstructions
	of the same signal (red dotted line) from data given in redshift space (grey
	dots) with Poissonian noise where the transformation from real to redshift
	space is given by \eqref{zdist:distortion-matrix}.
	The data is where the point-wise inversion from \eqref{mocksig:inverse-data}
	would place it.
	The smooth blue curve shows the classical signal reconstruction as proposed
	by our method with its $1\sigma$ error contours (thin blue lines).
	The bottom part of the nine panels show the characteristic residuals as
	defined by \eqref{def:characteristic-residual} for $500$ different data
	realisations of this signal.}
	\label{fig:ld-map0098}
\end{figure*}
We set up a run of 500 signal reconstructions on an interval of length $L=1$
with 1024 equally sized pixels.
Due to our limited interval length we had to restrict ourselves to signals whose
maximum peak was less than $3.5$, as for peaks higher than this value the tail
of the smeared-out peaks overlap with the tail on the other
side\footnote{As an estimate of how many signals this excludes, one can
calculate $P_{reject}\approx 1-\Bigl(\frac{1}{2}
\bigl(1+\mathrm{erf}\frac{3.5}{\sqrt{2}}\bigr)\Bigr)^{L/l_c}\approx
0.5\%$ which excludes only a minuscule subset of the possible
signals.}.

For each signal realisation $s$ we generate one possible data realisation $d$
and do three different types of reconstructions, namely the naive map \mnaive\
from \ref{mocksig:naive-map}, a MAP reconstruction neglecting redshift space
distortions \MAPu\ and the full MAP map including the model redshift space
distortions \MAP.
\figref{fig:ld-map0098} shows our \MAP\ reconstruction for one signal but for
different $b$ and $\bar{\rho}_{gal}$ settings, along with the characteristic
residuals $\res^{(\strue)}$ for the reconstruction of the signal in
the lower panels.
For our model we find the following general trends
\begin{itemize}
	\item the higher the galaxy density, the better the reconstruction
	\item the higher the bias, the worse the reconstruction of voids, but
	peaks on the other hand are better reconstructed
\end{itemize}
The reasons for these trends are the same as in section
\ref{section:ph:results}.
The reconstructions from galaxies with redshift distortions is therefore similar
in many respects to the one with photometric redshifts.

Note that the centre peak in the data (at $0.55$) shows a clear dislocation
from its signal counterpart in direction of the massive at $0.9$ -- an effect
due to the linear velocity distortions. 
The \MAP\ reconstruction however, being aware of the redshift distortions
introduced by the large massive, places the peak at the right spot in all
reconstructions.
Interestingly, for large biasses this peak is well resolved even for very low
galaxy densities as the characteristic residuals show.

Yet there are limitations for the reconstructions as can be seen in the
indentations flanking the largest peak at positions $0.8$ and $1.0$.
Although being comparable in size to the peak at $0.55$, they are not resolved
in any reconstruction, and this even remains the case for extremely large
$\bar{\rho}_{gal}$ which we do not show here.
Apparently, these structures are irreversibly lost in the shadow that the
larger peak at $0.9$ casts on close-by smaller structures.
This is also not unexpected, as the distortion matrix \mixingMatrix\ is set up
in a way that large peaks become smeared out over large distances, whereas small
peaks remain localised.
Therefore, the smaller peak appears as an extension of the plateau.
The most important characteristic of this kind of error is that it does not
improve with higher $\bar{\rho}_{gal}$ in contrast to areas where more
information can be gained by reducing the noise (e.g.~wide void regions).

\begin{figure}
	\includegraphics{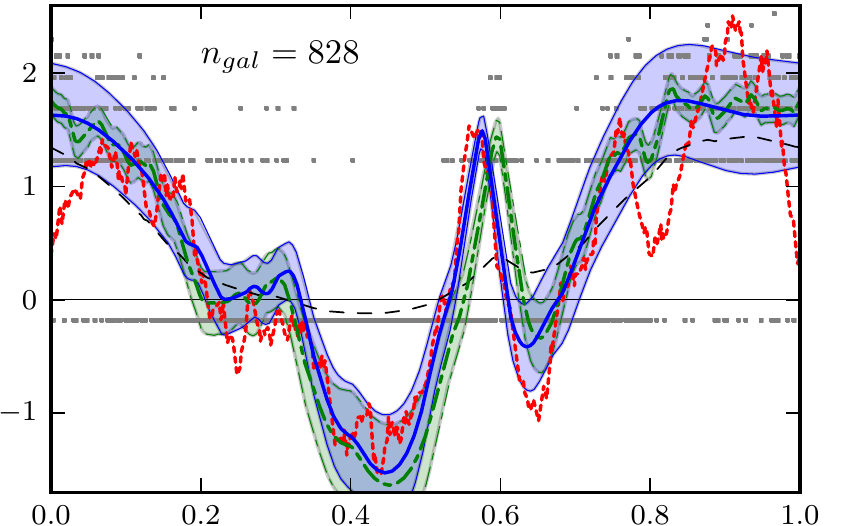}
	\caption[Reconstruction from a redshift distorted data set]%
	{Reconstruction of the same signal (red dotted line) as in
	\figref{fig:ld-map0098} for an average galaxy density $\bar{\rho}_{gal}=500$
	and galaxy bias $b=1.5$.
	The data (grey dots) is where the point-wise inversion from
	\eqref{mocksig:inverse-data} would place it and the thin evenly dashed black
	line shows the naïve map \mnaive.
	In addition to our redshift corrected map \MAP\ (blue smooth curve) we also
	show a MAP reconstruction \MAPu\ neglecting redshift distortions (green
	dash-dotted curve).
	Note that \MAPu\ and \mnaive\ pick up basically the same features, but
	\MAPu\ responds much quicker to an excess of galaxies.
	$1\sigma$ error levels are indicated by thin lines.}
	\label{fig:zdist:ld_bigmap0098}
\end{figure}
In \figurename~\ref{fig:zdist:ld_bigmap0098} we show the centre panel from
\figref{fig:ld-map0098} with the reconstruction for $b=1.5$ and
$\bar{\rho}_{gal}=500$ again but also the naive map \mnaive\ and the MAP map
neglecting redshift space distortions \MAPu.
At first sight, one notices the poor guess that \mnaive\ gives, which we
take as an argument that Bayesian analysis is inevitable to tackle this problem.

The more interesting competitor for the MAP map including the correction of
redshift space distortions \MAP\ is \MAPu.
In general, the shape of the reconstruction is very similar which is as it
should be, as both algorithms base on the same principles and work on the same
data set.
Yet there are some distinct features that make the difference.
Most prominent is of course the correct treatment of linear redshift space
distortions of \MAP\ in contrast to \MAPu\ which sees the void from $0.15$ to
$0.3$ further depleted and the peak at $0.55$ displaced towards the massive on
the right. 
Another more subtle difference is that \MAPu\ picks up more small scale features
from the data as can be seen in region $0.8$ to $1.1$.
This is due to the smoothing effect of $\mixingMatrix^\tp$ on the \MAP\
reconstruction as already discussed in section \ref{section:ph:results}.

In contrast to the photometric case from section \ref{section:ph:results}, the
error bars do not tighten up in overdense regions.
However this has to be expected, since our model was set up in a way, that the
position uncertainty in the neighbourhood of large peaks is largest.
In particular, this makes the detection of substructure in large peaks nearly
impossible.

\begin{table}
\centering
	\caption{%
		As an indicator for the quality of different reconstructions this
		table lists $\expectationWRT{\Lpnorm{m-\strue}{2}^2}{(\strue)}$, the
		average $L_2$-distance from reconstruction $m$ to true signal
		$\strue$, where the $s$-average runs over 500 random signal
		realisations.}
	\begin{tabular}{c|c||c|c|c}
		&& $\bar{\rho}_{gal} = 250$ & $\bar{\rho}_{gal} = 500$ & $\bar{\rho}_{gal} = 1000$ \\
		\hline
			& \mnaive &0.88 &0.81 & 0.71 \\
		$b=0.5$	& \MAPu &0.29 &0.24 & 0.21 \\
			& \MAP &0.26 &0.20 & 0.15 \\
		\hline
			& \mnaive &0.71 &0.62 & 0.54 \\
		$b=1.5$	& \MAPu &0.27 &0.23 & 0.21 \\
			& \MAP &0.23 &0.18 & 0.14 \\
		\hline
			& \mnaive &0.75 &0.70 & 0.66 \\
		$b=2.5$	& \MAPu &0.48 &0.42 & 0.38 \\
			& \MAP &0.41 &0.34 & 0.28
	\end{tabular}
	\label{tab:ld:avg-residuals}
\end{table}
For a signal-independent view on the reconstruction quality we list in
\tableref{tab:ld:avg-residuals} the average $L_2$-distance from
reconstruction to true signal for 500 different signal reconstructions.

The reconstruction benefit from including redshift space distortions seems not
to be overwhelming but tends to be larger if bias and galaxy density rise.
Notably for \MAPu\ the effects from linear and non-linear redshift distortions
partially cancel each other for the parameters chosen here and thereby improve
the performance of \MAPu.
This is because non-linear redshift space distortions smear out large peaks
while linear redshift distortions compress large overdensities.
If we turn off linear redshift distortions and set up \mixingMatrix\ only with
our virialisation model for non-linear redshift distortions, we get the
interesting effect that the average $L_2$-distance of \MAP\ to \strue\
becomes smaller, but for \MAPu\ it increases instead.

So far we have assumed that the underlying redshift distortion model is
perfectly known and its parameters are the same both for data generation and
reconstruction phase.
In a set-up for measured data this is not the case.
Neither is there a perfect model for the forward transformation from real to
redshift space, nor are its parameters accurately measured.
How severely parameter errors affect the reconstruction has not been
scrutinised, but can be aim of further investigation.
Still, if this model was adapted and applied to measured data, the above
mentioned reconstruction characteristics and limitations would hold.

\subsubsection{Refining the distortion
	model}\label{section:ld-distortion-model-refined}
In section \ref{section:ld-distortion-model} we have introduced a model for the
transformation from real to redshift space in a statistical way.
One important step was to apply a lowpass filter to the potential before the
linear velocity field was calculated.
However, similarly to using only the low modes of the perturbation for
calculating the velocity distortions, it would be reasonable to use only the
high modes for calculating the velocity dispersion.

\begin{figure}
	\begin{center}%
		\includegraphics{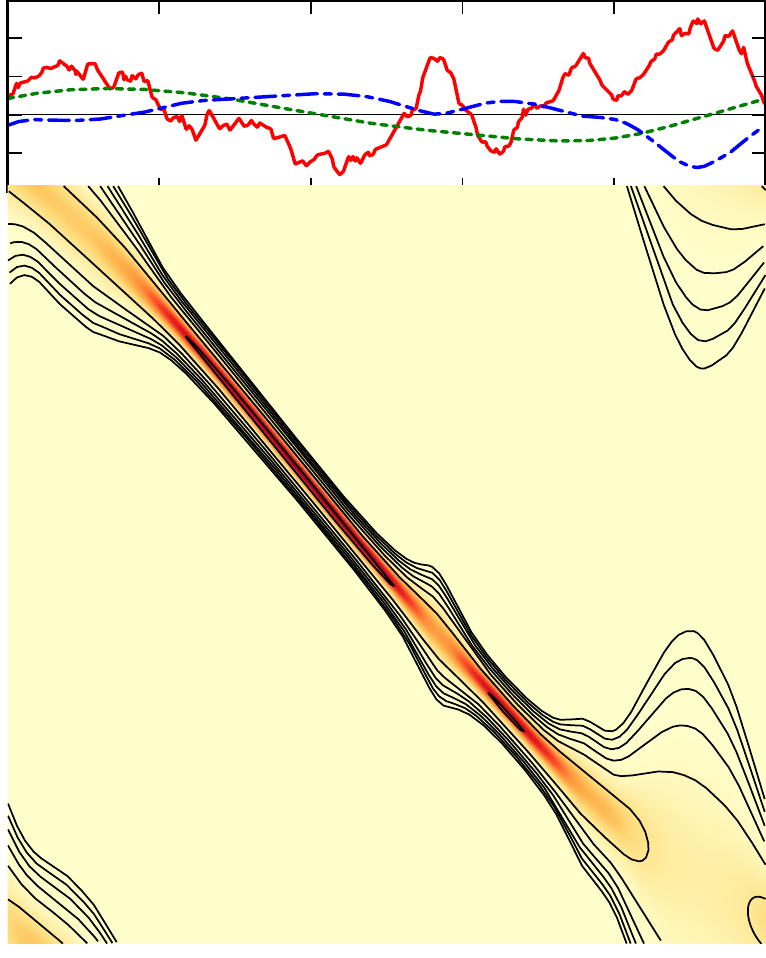}
	\end{center}
	\caption[Refined distortion matrix for transformation from real to redshift space]%
	{Example for a possible distortion matrix where the potential
	leading to the non-linear redshift distortions is high-pass filtered.
	Note that the maximum of the distribution is not necessarily on the diagonal
	due to the linear redshift distortions.
	The log-spaced contours indicate levels of equal height.
	On top the signal (solid red line) that generates this distortion matrix,
	the potential (blue dash-dotted line) and the streaming velocity resulting
	from linear theory (green dotted line) -- all scaled to comparable
	amplitudes.
	Note that in contrast to \figref{fig:zdist:distortionmatrix} the
	sharp peak at $0.55$ is now a separate collapsed object with its own
	velocity dispersion.}
	\label{fig:zdist:distortionmatrix-refined}
\end{figure}
This is because the non-linear velocity dispersion predominantly depend on
small-scale physics.
If a halo with many galaxies collapses, the energy that the galaxies gain from
the collapse is only the difference from their former potential energy and the
potential energy after they have entered the collapsed structure.
The velocity dispersion is hence not proportional to the full potential, but
only to the fraction of the potential that the galaxies have actually fallen.
Assuming that the galaxies in the collapsed structure are all from the vicinity
of the collapsed object, this fraction can be estimated as the high frequency
component of the potential.
In the picture of energy conservation one may look upon this as
\begin{itemize}
	\item the low frequency part of the potential adds to the energy of linear
	velocity distortions and
	\item the high frequency part of the potential adds to the non-linear
	velocity component.
\end{itemize}
In \figref{fig:zdist:distortionmatrix-refined} we show the resulting distortion
matrix of the modified model for the same signal as in
\figref{fig:zdist:distortionmatrix}.

When we set up the velocity distortions this way, we find that the problem
becomes severely harder and the MAP solution ultimately fails to give reasonable
results.
Interestingly it is the low bias case $b=0.5$ which is hardest.
This is most likely due to the fact that the response for $b=0.5$ makes a double
peak from large peaks, i.e.~bifurcates the peak.
This misleads the MAP map to make two peaks out of one, which can lead to very
weird behaviour.
This bifurcation will eventually also happen for the larger biasses, if we turn
up the strength of the velocity dispersion.

Our tests show, that the MAP solution via conjugate gradient minimization has
severe problems with this bifurcation.
We also employed a \emph{simulated annealing}\index{simulated annealing}
technique for minimization which gave us the same results.
So we can say with good confidence, that in the case of a bifurcated response
the MAP method may give the most likely map, but still a very bad
reconstruction.
With the complexity of the distortion matrix at this point we finally have
reached a limit for the MAP method.

\subsubsection{Comparison to Metropolis-Hastings sampling}
	\label{section:metropolis-hastings}
The question now arises, whether the MAP method is already the optimal
reconstruction given the data or not.
According to theory, this should not be the case, because the MAP map minimizes
$\expectationWRT{\Lpnorm{m-\strue}{0}}{(s|d)}$, but the $L_0$-norm is a rather
distorted measure of distance.
Unfortunately, the posterior is too complex that $\expectationWRT{s}{(s|d)}$ can
be calculated directly.
However, if there was a way to approximate the posterior by another \PDF, that
is easier to evaluate, we may be able to calculate $\expectationWRT{s}{(s|d)}$
directly.
Since our interest is not in the detailed shape of the posterior, but our aim is
simply to evaluate integrals over the posterior quickly, we may approximate the
posterior $P(s|d)$ by
\begin{equation}
	\Ptilde(s)\equiv \frac{1}{N} \sum_{i=1}^N\delta(s-s_i)
	\label{def:Ptilde}
\end{equation}
where we construct the $s_i$ using a Metropolis-Hastings MCMC method.
In principle the chain can start from any map $s_0$, but for the sake of
skipping the burn-in phase, we start our MCMC from \MAP.
At any given point $s_n$ we generate a small variation $\delta s$ with the same
power spectrum as the signal, but with far smaller amplitude and set
$s_{n+1}=s_n+\delta s$.
We accept or reject this new sample according to the Metropolis-Hastings
criterion \citep[][e.g.~]{1997RPPh...60..487B}.
This gives us a chain of maps where consecutive samples are correlated, but
after some steps the correlation vanishes.

\begin{figure*}
	\begin{center}%
		\includegraphics{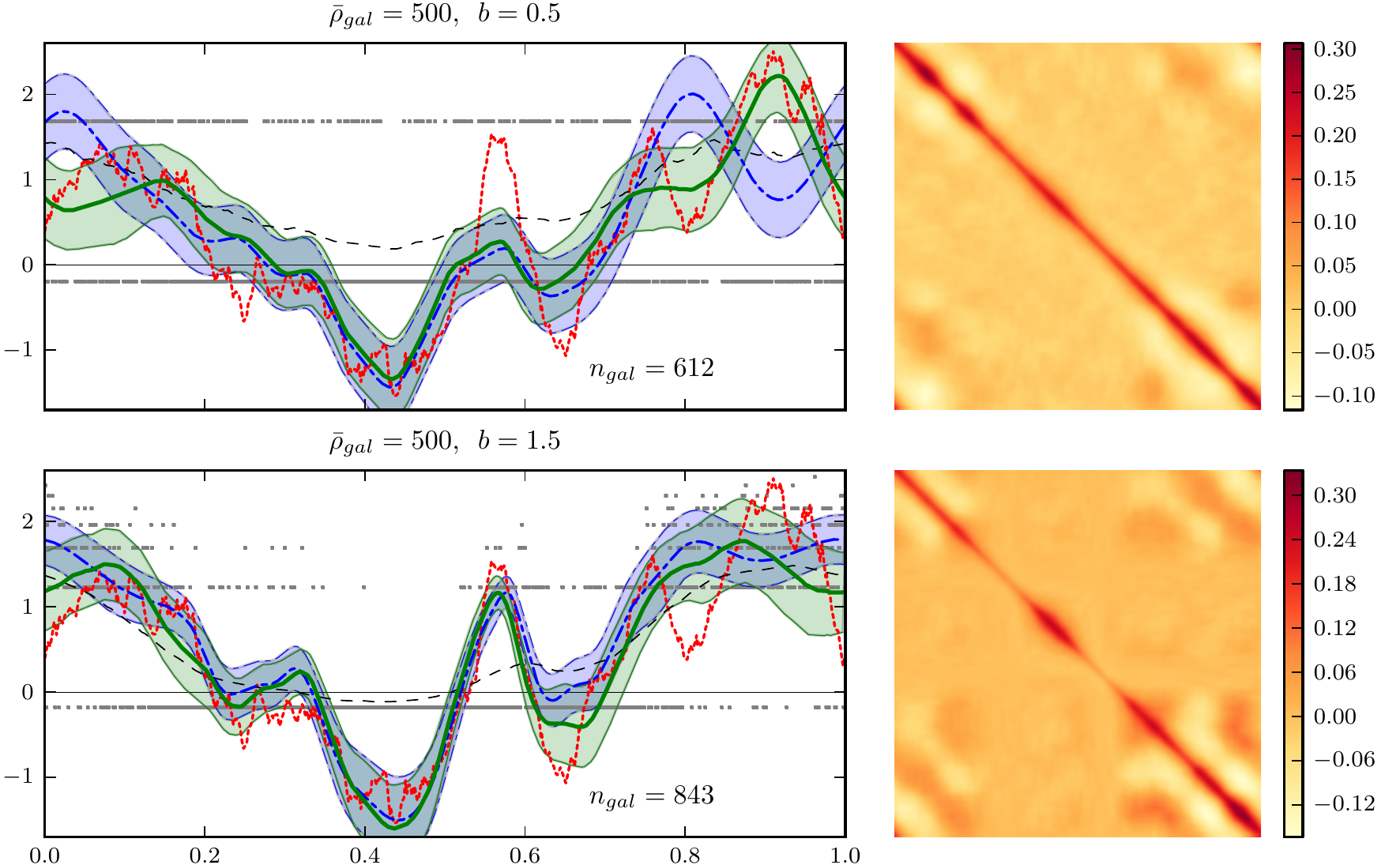}
	\end{center}
	\caption[Reconstruction using Metropolis-Hastings sampling]
		{On the left two reconstructions for different settings of the biasses
		$b=0.5$ and $b=1.5$.
		The signal appears as a red dotted line, the data (grey dots) are where
		the naïve data inversion from \eqref{mocksig:inverse-data} would place
		it.
		The naïve map (thin evenly dashed line) is the smoothed data and shows
		signs of bifurcation at $0.9$.
		The \MAP\ reconstruction (blue dash-dotted curve) interprets the
		bifurcation as two different peaks, while the posterior mean (smooth
		green curve) does not.
		$1\sigma$ error contours are indicated by thin lines.
		The pictures on the right show the uncertainty covariance matrix of the
		Metropolis-Hastings reconstruction.
		Note the long-range influence of the structure at $0.9$.}
	\label{fig:zdist:metropolis-hastings}
\end{figure*}
Since a MCMC is expensive \wrt\ computation time, we can not run several hundred
reconstructions but have to be satisfied with the evaluation of selected cases.
Therefore, we show in \figref{fig:zdist:metropolis-hastings} only three
reconstructions without the characteristic residuals as we did before.
As one can see there, the sampling method and the MAP method give very similar
results in the region from $0.2$ to $0.7$ where the redshift space distortions
are comparatively small, although the posterior mean tends to be a bit better
especially in the $b=1.5$ case.
Similarly, the width of the error bars is virtually the same in this region.

But not so in the region from $0.7$ to $1.2$\footnote{In order to not mix up the
ranges $[a,\,b]$ and $[b,\,1]\cup[0,\,a]$ we will say $[b,\,1+a]$ if we mean
the latter}.
There, the MAP approach is taken in by the bifurcation and gives a very bad
guess from $0.7$ to $1.1$ where it places peaks instead of valleys and a valley
instead of a peak.
And what makes the situation problematic indeed is that the estimated error bars
are rather small there.
The algorithm therefore completely misjudges its accuracy.
As already mentioned above, the situation improves a bit for the bias $1.5$ but
the problem is still there.

The sampling method on the other hand does the right thing in the interval $0.7$
to $1.1$.
It nicely reshapes the largest peak and at the place of the side-peak $0.75$
which is in the `shadow' of the larger peak at $0.9$ it widens the error bar
to remain on the safe side.
If this is just by chance of unlucky data cannot be judged for sure at this
point.
For the larger bias $1.5$ the improvement is not quite so good because the
shadow on nearby structures is much stronger.

The impression that the reconstruction quality of the average map from
Metropolis-Hastings sampling is much better than the \MAP\ reconstruction also
manifests itself in the $L_2$-distance which we list in
\tableref{tab:metropolis-hastings:L2-distances}.
\begin{table}
	\centering
	\caption{$L_2$-distance of reconstruction to signal for three reconstructions
		two of which are shown in \figref{fig:zdist:metropolis-hastings}.
		Note that there is hardly any improvement of the classical map from
		$\bar{\rho}_{gal} = 500$ to $\bar{\rho}_{gal} = 1000$ for $b=0.5$ while
		the posterior average improves by about 40\%.
		For the larger bias $b=1.5$ the difference between \MAP\ and posterior
		average becomes less palpably.}
	\begin{tabular}{r||c|c||c}
		& $\bar{\rho}_{gal} = 500$ & $\bar{\rho}_{gal} = 1000$ &
				$\bar{\rho}_{gal} = 500$ \\
		& $b=0.5$ & $b=0.5$ & $b=1.5$ \\
		\hline
		$\Lpnorm{\MAP-\strue}{2}^2$ \rule{0pt}{2.5ex}& 0.42 & 0.41 & 0.24 \\
		$\Lpnorm{\expectationWRT{s}{(s|d)}-\strue}{2}^2$ \rule{0pt}{2.5ex}& 0.17 & 0.10&0.16
	\end{tabular}
	\label{tab:metropolis-hastings:L2-distances}
\end{table}

Note also how the uncertainty covariance matrix in
\figref{fig:zdist:metropolis-hastings} changes from $b=0.5$ to $b=1.5$.
To understand what lies beneath one has to look at the Gaussian approximation of
the posterior $P(s|d)\approx \mathcal{N}\exp\bigl(j^\tp s -
\NormWithMetric{s}{\propD^{-1}}/2\bigr)$.
Here $\mathcal{N}=\e^{-\NormWithMetric{j}{\propD}/2}/\det{2\pi \propD}$ is a
normalisation factor and $j$ summarises all terms of the probability Hamiltonian
of first order in $s$.
Direct calculation yields
\begin{equation}
\expectationWRT{s}{(s|d)}\approx \mathcal{N}\PathInt{s}\,s\,\e^{j^\tp s -
		\NormWithMetric{s}{\propD^{-1}}/2} = \propD j
	\label{metropolis-hastings:gaussian-approximation}.
\end{equation}
\citet{2009PhRvD..80j5005E} call $j$ the \emph{information source}
\index{information source} because it excites the posterior mean from a zero-map
to a non-zero state.
One can show that in our problem $j$ contains a term linear in $d$ and
additional terms.
Therefore, equation \eqref{metropolis-hastings:gaussian-approximation} tells how
the posterior mean reacts on additional data.
Hence the pictures of the propagator in \figref{fig:zdist:metropolis-hastings}
show how the posterior map is constructed from the information source $j$ and
introduces a correlation length in the map.

In void regions the correlation length is longer than in overdense regions
because the Poissonian noise is larger there.
If there were no other source of uncertainty but the Poissonian noise, the band
would be tightest in the most overdense regions.
But there \textit{is} an additional source of uncertainty, namely the velocity
dispersion from the redshift distortion model.
Therefore, in overdense regions the correlation length of the propagator does
not collapse, but shows a pattern of strong correlation and anticorrelation,
as can be seen in the region from $0.7$ to $1.2$ in both cases.
Comparing the correlation length to the distortion matrix in
\figref{fig:zdist:distortionmatrix-refined} shows that this is in good agreement
to the correlation induced from the distortion matrix.
The main difference in the correlation matrix of the example with $b=1.5$ is
that the contrast is sharper and that the diagonal is much more structured.

\section{Conclusions}\label{section:conclusions}
Many reconstruction problems in cosmology suffer not only from large noise but
also from substantial measurement uncertainties.
While it is possible that some measurement uncertainties will be ameliorated in
the future by more sophisticated techniques, other sources of uncertainty are
fundamental such as cosmic variance and galaxy redshift distributions.
Areas where this applies are real space LSS reconstructions from galaxy counts
in redshift space, but also consistent treatment of photometric redshift.
For precision cosmology with galaxies it is therefore of paramount importance to
incorporate these uncertainties in the analysis.

Here we have presented a novel method how spatially distorted log-normal fields
as they occur in density field reconstruction can be reconstructed in a
Bayesian way.
This method was developed in the framework of information field theory which we
outlined in section \ref{section:theoretic-background}.
We showed that the IFT moment calculation ultimately foots on the minimization
of the expected $L_2$-weighted error of the reconstruction.
Where exact moment calculation from the posterior was not possible, we argued
how the correct map -- the posterior mean -- could be approximated by a
MAP approach.

We developed a data model for a log-normal signal with Poissonian noise where
the response can be non-local.
We even allowed for the case, in which the distortion of data space could depend
on the signal that was to be reconstructed.
The resulting problem is so complex that it could only be solved approximately
via numerical minimization of its probability Hamiltonian.

For a test of our approach we performed simulations where we constructed mock
signals, produced mock data thereof and tried to reconstruct the underlying
signal by numerical minimization of the probability Hamiltonian.

We tested our reconstruction code on three different distortion problems which
were
\begin{itemize}
	\item data with typical distortions as they appear in photometric redshift
	measurement,
	\item coded mask aperture problems as they appear in X- and $\gamma$-ray
	astronomy and
	\item real space matter reconstruction from redshift distorted data.
\end{itemize}
For the latter we developed a model for the forward problem to construct
redshift space data from real space galaxy distributions and where the
distortion was dependent on the underlying matter distribution that was to
be measured.
We were able to tackle this problem with a MAP approach.
However, after further complication of the distortion operator we found that the
MAP method does not live up to its expectations.
Instead, we could show that approximating the posterior via Metropolis-Hastings
sampling could give much more accurate reconstructions.
Therefore we think, that for such complicated problems the MAP method gives
misleading results and should be superseded by more powerful however also
computationally more demanding approaches such as sampling the posterior \PDF.

For the coded mask data we were able to identify the largest peaks and showed
that it is even possible to reconstruct their substructure if the count rates
are high enough.
An application of this approach to real X- or $\gamma$-ray data should be
possible but before doing so, some effort must be spent to make the approach
robuster to false detection of peaks.

At last, the reconstruction of a redshift space signal from photometric redshift
data proved to be very fruitful.
In many cases we were able to reconstruct the underlying matter distribution
remarkably well.
Since the colour space distortion is independent of the underlying signal, an
application of our approach to large data sets is feasible.

We also showed that in the IFT framework it is possible to easily combine data
sets with different error characteristics.
We considered the problem of combining photometric redshift data with large
uncertainties and spectroscopic data that are very accurate in position.
Our analysis showed that even with a low abundance of accurate data it is
possible to improve the reconstruction from distorted data with large abundance
as long as there is room for improvement.

In all cases we found that Bayesian analysis of the problem is inevitable for
the noise level we were considering.
We also showed that the reconstruction becomes significantly worse when the data
were distorted, but the data space distortion was neglected during the
reconstruction.
Therefore we think that including the data space distortions in future precision
analysis is inevitable.

Since the assumptions of our method are based on a few generic principles we are
confident that further areas will be found where our work will be appreciated.

\appendix
\section{Notation}\label{appendix:notation}
Having to deal extensively with field variables, one needs a shorthand notation
for the calculus.
Hence we define a vector space whose elements are fields -- in this sense we
use the terms `vector' and `field' interchangeably.
The dot product in this vector space is defined by
\begin{equation}
	v^\tp w\equiv \int dx\,v(x)w(x).
	\label{appendix:notation:vecproduct}
\end{equation}
Instead of writing the field variable in round brackets we usually use a
notation where the variable is in subscript to make the correspondence to finite
dimensional vector spaces more obvious.

We also use tensors of higher rank over this vector space, most notably
matrices.
Analogously to \eqref{appendix:notation:vecproduct}, we define
\begin{align*}
	\bigl(\mathbf{M}v\bigr)(x) &= \int dy\,M(x,\,y)v(y) \\
	\bigl(\mathbf{M}\mathbf{N}\bigr)(x,y) & = \int dz\,M(x,\,z)N(z,\,y)
\end{align*}
and so on.

We also need a field space equivalent of the frequently used rules
\begin{align*}
	\frac{\partial}{\partial x_i}x_j &= \delta_{ij} & \frac{\partial}{\partial
		x_i} \sum_jx_jk_j& =k_i.
\end{align*}
For all practical applications this is enough as one deals only with a finite
number of pixels, however all equations remain valid if one defines the
\emph{functional derivative}\index{functional derivative} according to
\citet[chap.~9]{1995iqft.book.....P} as
\begin{align*}
	\variation{}{J_x}J_y& =\delta(x-y)& \variation{}{J_x}\int dy\,J_y\Phi_y & =\Phi_x.
\end{align*}
This blends in naturally with our definition of the vector product
\eqref{appendix:notation:vecproduct}.

Vectors are printed in normal font with indices omitted.
Matrices are in bold font where possible, with the notable exception of
derivatives of vectors such as $\variation{\mu}{s}$.
To avoid an ambiguity in the sequence of the indices we define for derivations
that the new index shall be last, i.e.
\begin{equation*}
	\left(\variation{\mu}{s}\right)_{ij}\equiv \variation{\mu_i}{s_j}.
\end{equation*}

All functions are understood to act component-wise, that is $f(v)$ should be
read as $f(v)_i \equiv f(v_i)$.
In particular, this also applies to fundamental operations, such as
multiplication and division: $(v\cdot w)_i \equiv v_i\,w_i$ and $(v/w)\equiv
v_i/w_i$.
Note also that we do not adopt Einstein's summation convention, such that an
expression as $v_i\, \e^{w_i}$ should in fact be read as the number $v_i$
multiplied by the number $\e^{w_i}$.

In this notation subtle differences do matter, but it shortens the formulas
considerably.
Here some intricate examples:
\begin{gather*}
	\NormWithMetric{s}{\prop} = \sum_{i,j}\prop_{ij}\,s_i\,s_j = \prop_{ij}\,s^i\,s^j\\
	\bigl(v (\prop w)\bigr)_i = v_i\, \Bigl(\sum_j \prop_{ij}w_j\Bigr) = v_i\,\prop_{ij}w^j\\
	(v\,w^\tp)_{ij} = v_iw_j \\
	(v \prop)_{ij} = v_i \prop_{ij}\\
	(v^\tp \prop)_i = \sum_j v_j \prop_{ji} = v^j\prop_{ji}
\end{gather*}

\section[Generation of mock signals]%
	{Generating mock signals with Gaussian covariance
		matrix}\label{appendix:mocksig}
The aim is to generate signals with Gaussian covariance matrix
\begin{equation}
	\expectationWRT{ss^\tp}{(s)} = \prop
	\label{appendix:mocksig:problemdefinition}.
\end{equation}
By this definition, \prop\ is symmetric and positive definite such that it has a
root, i.e.~there exists a matrix \rprop\ such that $\prop = \rprop^\tp\rprop$.
We now show that the mock signal can be generated by convolving a vector of
Gaussian random numbers $r$ with \rprop: $s = \rprop r$

Without loss of generality we restrict our reasoning to the case where the
Gaussian random numbers in $r$ have unit variance, such that the prior for $r$
is
\begin{equation}
	P(r)=\prod_j \frac{1}{\sqrt{2\pi}}\e^{-r_j^2/2} =
		 \frac{\e^{-r^2/2}}{\det{2\pi\openone}^{1/2}}
	\label{appendix:mocksig:r-prior}.
\end{equation}
Here the product runs over all pixel indices.
From $r$ we construct the signal vector $s$ by application of
$\rprop$, which can be understood as a smoothing procedure for
the random values.
Note also that only in this step different points of the signal become
correlated with each other -- the entries of $r$ are by virtue of the random
nature of its entries uncorrelated.

We now have to prove that this procedure gives the desired prior for $s$:
\allowdisplaybreaks%
\begin{align*}
 P(s) & = \PathInt{r}\,\delta(s-\rprop r)\cdot P(r) \\
	& = \PathInt{q}\PathInt{r}\,\frac{1}{\det{2\pi
		\openone}}\e^{-iq^\tp(s-\rprop r)}\,
		\frac{\e^{-r^2/2}}{\det{2\pi\openone}^{1/2}}\\
	& = \frac{1}{\det{2\pi \openone}} \PathInt{q}\,
	\e^{-\NormWithMetric{q}{\prop}/2-\imath q^\tp s} \\
	& = \frac{1}{\det{2\pi \prop}^{1/2}}\,
		\e^{-\NormWithMetric{s}{\prop^{-1}}/2} =
		\gaussSymbol{s}{\prop}
\end{align*}

As source for random numbers we use the `\emph{Mersenne
Twister}'\index{Mersenne Twister} random number generator which offers a pseudo
random number sequence with a period of $2^{19937}-1$ \citep{272995}.
Its advantages are speed and very good randomness.
In particular, we use its implementation from the GNU Scientific Library
(gsl\footnote{The GNU Scientific Library is available from
\href{http://www.gnu.org/software/gsl}{http://www.gnu.org/software/gsl}}).

Note that this procedure gives an easy and robust way to test whether the
constructed signal has indeed the desired covariance
\begin{equation*}
	s^\tp\prop^{-1}s =
	r^\tp\rprop^\tp\prop^{-1}\rprop r =
	r^\tp r
	\approx N_{pix}
\label{appendix:mocksig:test-s}
\end{equation*}
where we have used that $r$ is a vector of $N_{pix}$ Gaussian random numbers of
unit variance in the last step.
In other words, for a Gaussian random field $s$ the number $s^\tp\prop^{-1}s$
should roughly give the dimension of $s$.

\bibliographystyle{mn2e}
\bibliography{mybibliography}

\begin{thebibliography}{}

\bibitem[\protect\citeauthoryear{{Ben{\'{\i}}tez}}{{Ben{\'{\i}}tez}}{2000}]{20%
00ApJ...536..571B}
{Ben{\'{\i}}tez} N.,  2000, \apj, 536, 571

\bibitem[\protect\citeauthoryear{{Bertschinger}, {Dekel}, {Faber}, {Dressler}
  \& {Burstein}}{{Bertschinger} et~al.}{1990}]{1990ApJ...364..370B}
{Bertschinger} E.,  {Dekel} A.,  {Faber} S.~M.,  {Dressler} A.,    {Burstein}
  D.,  1990, \apj, 364, 370

\bibitem[\protect\citeauthoryear{{Bialek}, {Callan} \& {Strong}}{{Bialek}
  et~al.}{1996}]{1996PhRvL..77.4693B}
{Bialek} W.,  {Callan} C.~G.,    {Strong} S.~P.,  1996, Physical Review
  Letters, 77, 4693

\bibitem[\protect\citeauthoryear{{Binder}}{{Binder}}{1997}]{1997RPPh...60..487%
B}
{Binder} K.,  1997, Reports on Progress in Physics, 60, 487

\bibitem[\protect\citeauthoryear{{Branchini}, {Plionis} \&
  {Sciama}}{{Branchini} et~al.}{1996}]{1996ApJ...461L..17B}
{Branchini} E.,  {Plionis} M.,    {Sciama} D.~W.,  1996, \apjl, 461, L17

\bibitem[\protect\citeauthoryear{{Bromley}, {Warren} \& {Zurek}}{{Bromley}
  et~al.}{1997}]{1997ApJ...475..414B}
{Bromley} B.~C.,  {Warren} M.~S.,    {Zurek} W.~H.,  1997, \apj, 475, 414

\bibitem[\protect\citeauthoryear{{Cole}, {Lacey}, {Baugh} \& {Frenk}}{{Cole}
  et~al.}{2000}]{2000MNRAS.319..168C}
{Cole} S.,  {Lacey} C.~G.,  {Baugh} C.~M.,    {Frenk} C.~S.,  2000, \mnras,
  319, 168

\bibitem[\protect\citeauthoryear{{Coles} \& {Jones}}{{Coles} \&
  {Jones}}{1991}]{1991MNRAS.248....1C}
{Coles} P.,  {Jones} B.,  1991, \mnras, 248, 1

\bibitem[\protect\citeauthoryear{{Cornwell} \& {Evans}}{{Cornwell} \&
  {Evans}}{1985}]{1985A&A...143...77C}
{Cornwell} T.~J.,  {Evans} K.~F.,  1985, \aap, 143, 77

\bibitem[\protect\citeauthoryear{{Dicke}}{{Dicke}}{1968}]{1968ApJ...153L.101D}
{Dicke} R.~H.,  1968, \apjl, 153, L101

\bibitem[\protect\citeauthoryear{{Duane}, {Kennedy}, {Pendleton} \&
  {Roweth}}{{Duane} et~al.}{1987}]{1987PhLB..195..216D}
{Duane} S.,  {Kennedy} A.~D.,  {Pendleton} B.~J.,    {Roweth} D.,  1987,
  Physics Letters B, 195, 216

\bibitem[\protect\citeauthoryear{{Efstathiou}}{{Efstathiou}}{2000}]{2000MNRAS.%
317..697E}
{Efstathiou} G.,  2000, \mnras, 317, 697

\bibitem[\protect\citeauthoryear{{En{\ss}lin}, {Frommert} \&
  {Kitaura}}{{En{\ss}lin} et~al.}{2009}]{2009PhRvD..80j5005E}
{En{\ss}lin} T.~A.,  {Frommert} M.,    {Kitaura} F.~S.,  2009, \prd, 80, 105005

\bibitem[\protect\citeauthoryear{{Enßlin} \& {Frommert}}{{Enßlin} \&
  {Frommert}}{2010}]{2010arXiv1002.2928E}
{Enßlin} T.,  {Frommert} M.,  2010, ArXiv e-prints

\bibitem[\protect\citeauthoryear{{Erdo{\u g}du}, {Lahav}, {Huchra} \& {et
  al.}}{{Erdo{\u g}du} et~al.}{2006}]{2006MNRAS.373...45E}
{Erdo{\u g}du} P.,  {Lahav} O.,  {Huchra} J.,    {et al.} 2006, \mnras, 373, 45

\bibitem[\protect\citeauthoryear{Erdo{\u g}du et al.}{2004}]{2004MNRAS.352..939E} Erdo{\u g}du P., et al., 2004, MNRAS, 352, 939

\bibitem[\protect\citeauthoryear{{Frieden} \& {Wells}}{{Frieden} \&
  {Wells}}{1978}]{1978JOSA...68...93F}
{Frieden} R.~R.,  {Wells} D.~C.,  1978, Journal of the Optical Society of
  America (1917-1983), 68, 93

\bibitem[\protect\citeauthoryear{{Fry} \& {Gaztanaga}}{{Fry} \&
  {Gaztanaga}}{1993}]{1993ApJ...413..447F}
{Fry} J.~N.,  {Gaztanaga} E.,  1993, \apj, 413, 447

\bibitem[\protect\citeauthoryear{Geman \& Geman}{Geman \&
  Geman}{1984}]{citeulike:893708}
Geman S.,  Geman D.,  1984, IEEE Trans. Pattern Anal. Mach. Intell., 6, 721

\bibitem[\protect\citeauthoryear{{Goldberg}}{{Goldberg}}{2001}]{2001ApJ...550.%
..87G}
{Goldberg} D.~M.,  2001, \apj, 550, 87

\bibitem[\protect\citeauthoryear{Green}{Green}{1990}]{Green90bayesianreconstru%
ctions}
Green P.~J.,  1990, IEEE Trans. Med. Imag, pp 84--93

\bibitem[\protect\citeauthoryear{{Gull} \& {Daniell}}{{Gull} \&
  {Daniell}}{1978}]{1978Natur.272..686G}
{Gull} S.~F.,  {Daniell} G.~J.,  1978, \nat, 272, 686

\bibitem[\protect\citeauthoryear{{Hamilton}}{{Hamilton}}{1995}]{1995clun.conf.%
.143H}
{Hamilton} A.~J.~S.,  1995, in {S.~Maurogordato, C.~Balkowski, C.~Tao, \&
  J.~Tran Thanh van} ed., Clustering in the Universe {Redshift distortions and
  {$\Omega$} in IRAS Surveys}.
p.~143

\bibitem[\protect\citeauthoryear{{Hamilton}}{{Hamilton}}{1998}]{1998ASSL..231.%
.185H}
{Hamilton} A.~J.~S.,  1998, in {D.~Hamilton} ed., The Evolving Universe
  Vol.~231 of Astrophysics and Space Science Library, {Linear Redshift
  Distortions: a Review}.
p.~185

\bibitem[\protect\citeauthoryear{Hastings}{Hastings}{1970}]{citeulike:1015842}
Hastings W.~K.,  1970, Biometrika, 57, 97

\bibitem[\protect\citeauthoryear{{Heavens} \& {Taylor}}{{Heavens} \&
  {Taylor}}{1995}]{1995MNRAS.275..483H}
{Heavens} A.~F.,  {Taylor} A.~N.,  1995, \mnras, 275, 483

\bibitem[\protect\citeauthoryear{{Hebert} \& {Leahy}}{{Hebert} \&
  {Leahy}}{1992}]{1992ITSP...40.2290H}
{Hebert} T.~J.,  {Leahy} R.,  1992, IEEE Transactions on Signal Processing, 40,
  2290

\bibitem[\protect\citeauthoryear{{Hoffman} \& {Zaroubi}}{{Hoffman} \&
  {Zaroubi}}{2000}]{2000ApJ...535L...5H}
{Hoffman} Y.,  {Zaroubi} S.,  2000, \apjl, 535, L5

\bibitem[\protect\citeauthoryear{{Hubble}}{{Hubble}}{1934}]{1934ApJ....79....8%
H}
{Hubble} E.,  1934, \apj, 79, 8

\bibitem[\protect\citeauthoryear{{Huchra}, {Jarrett}, {Skrutskie}, {Cutri},
  {Schneider}, {Macri}, {Steining}, {Mader}, {Martimbeau} \& {George}}{{Huchra}
  et~al.}{2005}]{2005ASPC..329..135H}
{Huchra} J.,  {Jarrett} T.,  {Skrutskie} M.,  {Cutri} R.,  {Schneider} S.,
  {Macri} L.,  {Steining} R.,  {Mader} J.,  {Martimbeau} N.,    {George} T.,
  2005, in {Fairall} A.~P.,  {Woudt} P.~A.,  eds, ASP Conf. Ser. 329: Nearby
  Large-Scale Structures and the Zone of Avoidance {The 2MASS Redshift Survey
  and Low Galactic Latitude Large-Scale Structure}.
p.~135

\bibitem[\protect\citeauthoryear{{Jasche} \& {Kitaura}}{{Jasche} \&
  {Kitaura}}{2009}]{2009arXiv0911.2496J}
{Jasche} J.,  {Kitaura} F.~S.,  2009, ArXiv e-prints

\bibitem[\protect\citeauthoryear{{Jasche}, {Kitaura}, {Li} \&
  {Enßlin}}{{Jasche} et~al.}{2009}]{2009arXiv0911.2498J}
{Jasche} J.,  {Kitaura} F.~S.,  {Li} C.,    {Enßlin} T.~A.,  2009, ArXiv
  e-prints

\bibitem[\protect\citeauthoryear{Jaynes}{Jaynes}{1957}]{PhysRev.106.620}
Jaynes E.~T.,  1957, Phys. Rev., 106, 620

\bibitem[\protect\citeauthoryear{{Jenkins}, {Frenk}, {White}, {Colberg},
  {Cole}, {Evrard}, {Couchman} \& {Yoshida}}{{Jenkins}
  et~al.}{2001}]{2001MNRAS.321..372J}
{Jenkins} A.,  {Frenk} C.~S.,  {White} S.~D.~M.,  {Colberg} J.~M.,  {Cole} S.,
  {Evrard} A.~E.,  {Couchman} H.~M.~P.,    {Yoshida} N.,  2001, \mnras, 321,
  372

\bibitem[\protect\citeauthoryear{{Jeong} \& {Komatsu}}{{Jeong} \&
  {Komatsu}}{2009}]{2009ApJ...703.1230J}
{Jeong} D.,  {Komatsu} E.,  2009, \apj, 703, 1230

\bibitem[\protect\citeauthoryear{{Kaiser}}{{Kaiser}}{1987}]{1987MNRAS.227....1%
K}
{Kaiser} N.,  1987, \mnras, 227, 1

\bibitem[\protect\citeauthoryear{{Kauffmann}, {White} \&
  {Guiderdoni}}{{Kauffmann} et~al.}{1993}]{1993MNRAS.264..201K}
{Kauffmann} G.,  {White} S.~D.~M.,    {Guiderdoni} B.,  1993, \mnras, 264, 201

\bibitem[\protect\citeauthoryear{{Kayo}, {Taruya} \& {Suto}}{{Kayo}
  et~al.}{2001}]{2001ApJ...561...22K}
{Kayo} I.,  {Taruya} A.,    {Suto} Y.,  2001, \apj, 561, 22

\bibitem[\protect\citeauthoryear{{Kitaura}, {Jasche} \& {Metcalf}}{{Kitaura}
  et~al.}{2010}]{2010MNRAS.403..589K}
{Kitaura} F.,  {Jasche} J.,    {Metcalf} R.~B.,  2010, \mnras, 403, 589

\bibitem[\protect\citeauthoryear{{Kitaura}, {Jasche}, {Li}, {En{\ss}lin},
  {Metcalf}, {Wandelt}, {Lemson} \& {White}}{{Kitaura}
  et~al.}{2009}]{2009MNRAS.400..183K}
{Kitaura} F.~S.,  {Jasche} J.,  {Li} C.,  {En{\ss}lin} T.~A.,  {Metcalf} R.~B.,
   {Wandelt} B.~D.,  {Lemson} G.,    {White} S.~D.~M.,  2009, \mnras, 400, 183

\bibitem[\protect\citeauthoryear{{Landau} \& {Lifshitz}}{{Landau} \&
  {Lifshitz}}{1966}]{1966mech.book.....L}
{Landau} L.~D.,  {Lifshitz} E.~M.,  1966, {Mechanik}

\bibitem[\protect\citeauthoryear{{Layzer}}{{Layzer}}{1956}]{1956AJ.....61..383%
L}
{Layzer} D.,  1956, \aj, 61, 383

\bibitem[\protect\citeauthoryear{{Lemm}}{{Lemm}}{1999}]{1999physics..12005L}
{Lemm} J.~C.,  1999, ArXiv Physics e-prints

\bibitem[\protect\citeauthoryear{{Lemm}}{{Lemm}}{2001}]{2001AIPC..568..425L}
{Lemm} J.~C.,  2001, in {A.~Mohammad-Djafari} ed., Bayesian Inference and
  Maximum Entropy Methods in Science and Engineering Vol.~568 of American
  Institute of Physics Conference Series, {Bayesian field theory and
  approximate symmetries}.
pp 425--436

\bibitem[\protect\citeauthoryear{{Lemm}, {Uhlig} \& {Weiguny}}{{Lemm}
  et~al.}{2001}]{2001EPJB...20..349L}
{Lemm} J.~C.,  {Uhlig} J.,    {Weiguny} A.,  2001, European Physical Journal B,
  20, 349

\bibitem[\protect\citeauthoryear{{Lemm}, {Uhlig} \& {Weiguny}}{{Lemm}
  et~al.}{2005}]{2005EPJB...46...41L}
{Lemm} J.~C.,  {Uhlig} J.,    {Weiguny} A.,  2005, European Physical Journal B,
  46, 41

\bibitem[\protect\citeauthoryear{{Mart{\'{\i}}nez} \& {Saar}}{{Mart{\'{\i}}nez}
  \& {Saar}}{2002}]{2002sgd..book.....M}
{Mart{\'{\i}}nez} V.~J.,  {Saar} E.,  2002, {Statistics of the Galaxy
  Distribution}.
Chapman \&amp

\bibitem[\protect\citeauthoryear{{Matsubara}}{{Matsubara}}{2008a}]{2008PhRvD..%
78j9901M}
{Matsubara} T.,  2008a, \prd, 78, 109901

\bibitem[\protect\citeauthoryear{{Matsubara}}{{Matsubara}}{2008b}]{2008PhRvD..%
78h3519M}
{Matsubara} T.,  2008b, \prd, 78, 083519

\bibitem[\protect\citeauthoryear{Matsumoto \& Nishimura}{Matsumoto \&
  Nishimura}{1998}]{272995}
Matsumoto M.,  Nishimura T.,  1998, ACM Trans. Model. Comput. Simul., 8, 3

\bibitem[\protect\citeauthoryear{{Metropolis}, {Rosenbluth}, {Rosenbluth},
  {Teller} \& {Teller}}{{Metropolis} et~al.}{1953}]{1953JChPh..21.1087M}
{Metropolis} N.,  {Rosenbluth} A.~W.,  {Rosenbluth} M.~N.,  {Teller} A.~H.,
  {Teller} E.,  1953, \jcp, 21, 1087

\bibitem[\protect\citeauthoryear{{Mukhanov}}{{Mukhanov}}{2005}]{2005pfc..book.%
....M}
{Mukhanov} V.,  2005, {Physical Foundations of Cosmology}

\bibitem[\protect\citeauthoryear{{Navarro}, {Frenk} \& {White}}{{Navarro}
  et~al.}{1996}]{1996ApJ...462..563N}
{Navarro} J.~F.,  {Frenk} C.~S.,    {White} S.~D.~M.,  1996, \apj, 462, 563

\bibitem[\protect\citeauthoryear{{Neyrinck}, {Szapudi} \& {Szalay}}{{Neyrinck}
  et~al.}{2009}]{2009ApJ...698L..90N}
{Neyrinck} M.~C.,  {Szapudi} I.,    {Szalay} A.~S.,  2009, \apjl, 698, L90

\bibitem[\protect\citeauthoryear{{Nityananda} \& {Narayan}}{{Nityananda} \&
  {Narayan}}{1982}]{1982JApA....3..419N}
{Nityananda} R.,  {Narayan} R.,  1982, Journal of Astrophysics and Astronomy,
  3, 419

\bibitem[\protect\citeauthoryear{{Nunez} \& {Llacer}}{{Nunez} \&
  {Llacer}}{1990}]{1990Ap&SS.171..341N}
{Nunez} J.,  {Llacer} J.,  1990, \apss, 171, 341

\bibitem[\protect\citeauthoryear{{Nusser} \& {Haehnelt}}{{Nusser} \&
  {Haehnelt}}{1999}]{1999MNRAS.303..179N}
{Nusser} A.,  {Haehnelt} M.,  1999, \mnras, 303, 179

\bibitem[\protect\citeauthoryear{{Oh} \& {Frieden}}{{Oh} \&
  {Frieden}}{2009}]{2009OptCo.282.2489O}
{Oh} C.,  {Frieden} R.~B.,  2009, Optics Communications, 282, 2489

\bibitem[\protect\citeauthoryear{{Peacock} \& {Dodds}}{{Peacock} \&
  {Dodds}}{1994}]{1994MNRAS.267.1020P}
{Peacock} J.~A.,  {Dodds} S.~J.,  1994, \mnras, 267, 1020

\bibitem[\protect\citeauthoryear{{Peebles}}{{Peebles}}{1980}]{1980lssu.book...%
..P}
{Peebles} P.~J.~E.,  1980, {The large-scale structure of the universe}

\bibitem[\protect\citeauthoryear{{Percival}}{{Percival}}{2005}]{2005MNRAS.356.%
1168P}
{Percival} W.~J.,  2005, \mnras, 356, 1168

\bibitem[\protect\citeauthoryear{{Percival}, {Nichol}, {Eisenstein}, {Frieman},
  {Fukugita}, {Loveday}, {Pope}, {Schneider}, {Szalay}, {Tegmark}, {Vogeley},
  {Weinberg}, {Zehavi}, {Bahcall}, {Brinkmann}, {Connolly} \&
  {Meiksin}}{{Percival} et~al.}{2007}]{2007ApJ...657..645P}
{Percival} W.~J.,  {Nichol} R.~C.,  {Eisenstein} D.~J.,  {Frieman} J.~A.,
  {Fukugita} M.,  {Loveday} J.,  {Pope} A.~C.,  {Schneider} D.~P.,  {Szalay}
  A.~S.,  {Tegmark} M.,  {Vogeley} M.~S.,  {Weinberg} D.~H.,  {Zehavi} I.,
  {Bahcall} N.~A.,  {Brinkmann} J.,  {Connolly} A.~J.,    {Meiksin} A.,  2007,
  \apj, 657, 645

\bibitem[\protect\citeauthoryear{{Peskin} \& {Schroeder}}{{Peskin} \&
  {Schroeder}}{1995}]{1995iqft.book.....P}
{Peskin} M.~E.,  {Schroeder} D.~V.,  1995, {An Introduction to Quantum Field
  Theory}.
Westview Press

\bibitem[\protect\citeauthoryear{{Ratcliffe}, {Shanks}, {Parker} \&
  {Fong}}{{Ratcliffe} et~al.}{1998}]{1998MNRAS.296..191R}
{Ratcliffe} A.,  {Shanks} T.,  {Parker} Q.~A.,    {Fong} R.,  1998, \mnras,
  296, 191

\bibitem[\protect\citeauthoryear{{Schmoldt}, {Saar}, {Saha}, {Branchini},
  {Efstathiou}, {Frenk}, {Keeble}, {Maddox}, {McMahon}, {Oliver},
  {Rowan-Robinson}, {Saunders}, {Sutherland}, {Tadros} \& {White}}{{Schmoldt}
  et~al.}{1999}]{1999AJ....118.1146S}
{Schmoldt} I.~M.,  {Saar} V.,  {Saha} P.,  {Branchini} E.,  {Efstathiou} G.~P.,
   {Frenk} C.~S.,  {Keeble} O.,  {Maddox} S.,  {McMahon} R.,  {Oliver} S.,
  {Rowan-Robinson} M.,  {Saunders} W.,  {Sutherland} W.~J.,  {Tadros} H.,
  {White} S.~D.~M.,  1999, \apj, 118, 1146

\bibitem[\protect\citeauthoryear{{Shaya}, {Peebles} \& {Tully}}{{Shaya}
  et~al.}{1995}]{1995ApJ...454...15S}
{Shaya} E.~J.,  {Peebles} P.~J.~E.,    {Tully} R.~B.,  1995, \apj, 454, 15

\bibitem[\protect\citeauthoryear{{Shewchuk}}{{Shewchuk}}{1994}]{shewchuk}
{Shewchuk} J.~R.,  1994, {An Introduction to the Conjugate Gradient Method
  Without the Agonizing Pain}.
published in the web

\bibitem[\protect\citeauthoryear{{Smith}, {Peacock}, {Jenkins}, {White},
  {Frenk}, {Pearce}, {Thomas}, {Efstathiou} \& {Couchman}}{{Smith}
  et~al.}{2003}]{2003MNRAS.341.1311S}
{Smith} R.~E.,  {Peacock} J.~A.,  {Jenkins} A.,  {White} S.~D.~M.,  {Frenk}
  C.~S.,  {Pearce} F.~R.,  {Thomas} P.~A.,  {Efstathiou} G.,    {Couchman}
  H.~M.~P.,  2003, \mnras, 341, 1311

\bibitem[\protect\citeauthoryear{{Springel}, {White}, {Jenkins}, {Frenk},
  {Yoshida}, {Gao}, {Navarro}, {Thacker}, {Croton}, {Helly}, {Peacock}, {Cole},
  {Thomas}, {Couchman}, {Evrard}, {Colberg} \& {Pearce}}{{Springel}
  et~al.}{2005}]{2005Natur.435..629S}
{Springel} V.,  {White} S.~D.~M.,  {Jenkins} A.,  {Frenk} C.~S.,  {Yoshida} N.,
   {Gao} L.,  {Navarro} J.,  {Thacker} R.,  {Croton} D.,  {Helly} J.,
  {Peacock} J.~A.,  {Cole} S.,  {Thomas} P.,  {Couchman} H.,  {Evrard} A.,
  {Colberg} J.,    {Pearce} F.,  2005, \nat, 435, 629

\bibitem[\protect\citeauthoryear{{Tadros}, {Ballinger}, {Taylor}, {Heavens},
  {Efstathiou}, {Saunders}, {Frenk}, {Keeble}, {McMahon}, {Maddox}, {Oliver},
  {Rowan-Robinson}, {Sutherland} \& {White}}{{Tadros}
  et~al.}{1999}]{1999MNRAS.305..527T}
{Tadros} H.,  {Ballinger} W.~E.,  {Taylor} A.~N.,  {Heavens} A.~F.,
  {Efstathiou} G.,  {Saunders} W.,  {Frenk} C.~S.,  {Keeble} O.,  {McMahon} R.,
   {Maddox} S.~J.,  {Oliver} S.,  {Rowan-Robinson} M.,  {Sutherland} W.~J.,
  {White} S.~D.~M.,  1999, \mnras, 305, 527

\bibitem[\protect\citeauthoryear{{Tadros} \& {Efstathiou}}{{Tadros} \&
  {Efstathiou}}{1996}]{1996MNRAS.282.1381T}
{Tadros} H.,  {Efstathiou} G.,  1996, \mnras, 282, 1381

\bibitem[\protect\citeauthoryear{{Vogeley}, {Hoyle}, {Rojas} \&
  {Goldberg}}{{Vogeley} et~al.}{2004}]{2004ogci.conf....5V}
{Vogeley} M.~S.,  {Hoyle} F.,  {Rojas} R.~R.,    {Goldberg} D.~M.,  2004, in
  {Diaferio} A.,  ed., IAU Colloq. 195: Outskirts of Galaxy Clusters: Intense
  Life in the Suburbs {Mapping the cosmic web with the Sloan Digital Sky
  Survey}.
pp 5--11

\bibitem[\protect\citeauthoryear{{Wang}, {Fu} \& {Qi}}{{Wang}
  et~al.}{2008}]{2008PMB....53..593W}
{Wang} G.,  {Fu} L.,    {Qi} J.,  2008, Physics in Medicine and Biology, 53,
  593

\bibitem[\protect\citeauthoryear{{Webster}, {Lahav} \& {Fisher}}{{Webster}
  et~al.}{1997}]{1997MNRAS.287..425W}
{Webster} M.,  {Lahav} O.,    {Fisher} K.,  1997, \mnras, 287, 425

\bibitem[\protect\citeauthoryear{{White} \& {Frenk}}{{White} \&
  {Frenk}}{1991}]{1991ApJ...379...52W}
{White} S.~D.~M.,  {Frenk} C.~S.,  1991, \apj, 379, 52

\bibitem[\protect\citeauthoryear{{Wittman}}{{Wittman}}{2009}]{2009ApJ...700L.1%
74W}
{Wittman} D.,  2009, \apjl, 700, L174

\bibitem[\protect\citeauthoryear{{Yahil}, {Strauss}, {Davis} \&
  {Huchra}}{{Yahil} et~al.}{1991}]{1991ApJ...372..380Y}
{Yahil} A.,  {Strauss} M.~A.,  {Davis} M.,    {Huchra} J.~P.,  1991, \apj, 372,
  380

\end{thebibliography}

\end{document}